\documentclass[manuscript]{acmart}

\usepackage{changepage}
\newcommand{\quotateblock}[2]{%
\vspace{0.3em}
\begin{adjustwidth}{1.0em}{0.5em}%
#1:\textit{``#2'’}%
\end{adjustwidth}%
\vspace{0.3em}
}

\newcommand\red[1]{{\color{black}#1}}

\usepackage{outlines}
\AtBeginDocument{%
  \providecommand\BibTeX{{%
    \normalfont B\kern-0.5em{\scshape i\kern-0.25em b}\kern-0.8em\TeX}}}

\begin{document}

\title{Towards Automated Accessibility Report Generation for Mobile Apps}

\author{Amanda Swearngin}
\email{aswearngin@apple.com}
\affiliation{
    \institution{Apple} 
    \country{USA}
}

\author{Jason Wu}
\email{jsonwu@cmu.edu}
\authornote{Work done while Jason Wu was an intern at Apple}
\affiliation{%
  \institution{HCI Institute, Carnegie Mellon University}
  \city{Pittsburgh}
  \state{PA}
  \country{USA}
}

\author{Xiaoyi Zhang}
\email{xiaoyiz@apple.com}
\affiliation{
    \institution{Apple} 
    \country{USA}
}

\author{Esteban Gomez}
\email{estebangomez@apple.com}
\affiliation{
    \institution{Apple} 
    \country{USA}
}

\author{Jen Coughenour}
\email{jjco@apple.com}
\affiliation{
    \institution{Apple} 
    \country{USA}
}

\author{Rachel Stukenborg}
\email{rstukenborg@apple.com}
\affiliation{
    \institution{Apple} 
    \country{USA}
}

\author{Bhavya Garg}
\email{bgarg@apple.com}
\affiliation{
    \institution{Apple} 
    \country{USA}
}

\author{Greg Hughes}
\affiliation{
    \institution{Apple} 
    \country{USA}
}

\author{Adriana Hilliard}
\email{adri@apple.com}
\affiliation{
    \institution{Apple} 
    \country{USA}
}

\author{Jeffrey P. Bigham}
\email{adri@apple.com}
\affiliation{
    \institution{Apple} 
    \country{USA}
}

\author{Jeffrey Nichols}
\email{jwnichols@apple.com}
\affiliation{
    \institution{Apple} 
    \country{USA}
}

\renewcommand{\shortauthors}{Swearngin, Wu, Zhang, et al.}

\begin{abstract}
Many apps have basic accessibility issues, like missing labels or low contrast. Automated tools can help app developers catch basic issues, but can be laborious or require writing dedicated tests. We propose a system, motivated by a collaborative process with accessibility stakeholders at a large technology company, to generate whole app accessibility reports by combining varied data collection methods (e.g., app crawling, manual recording) with an existing accessibility scanner. Many such scanners are based on single-screen scanning, and a key problem in whole app accessibility reporting is to effectively de-duplicate and summarize issues collected across an app. To this end, we developed a screen grouping model with 96.9\% accuracy (88.8\% F1-score) and UI element matching heuristics with 97\% accuracy (98.2\% F1-score). We combine these technologies in a system to report and summarize unique issues across an app, and enable a unique pixel-based ignore feature to help engineers and testers better manage reported issues across their app's lifetime. We conducted a qualitative evaluation with 18 accessibility-focused engineers and testers which showed this system can enhance their existing accessibility testing toolkit and address key limitations in current accessibility scanning tools.
\end{abstract}

\keywords{user interface modeling, app crawling, accessibility}

\begin{teaserfigure}
    \centering
    \includegraphics[width=\textwidth]{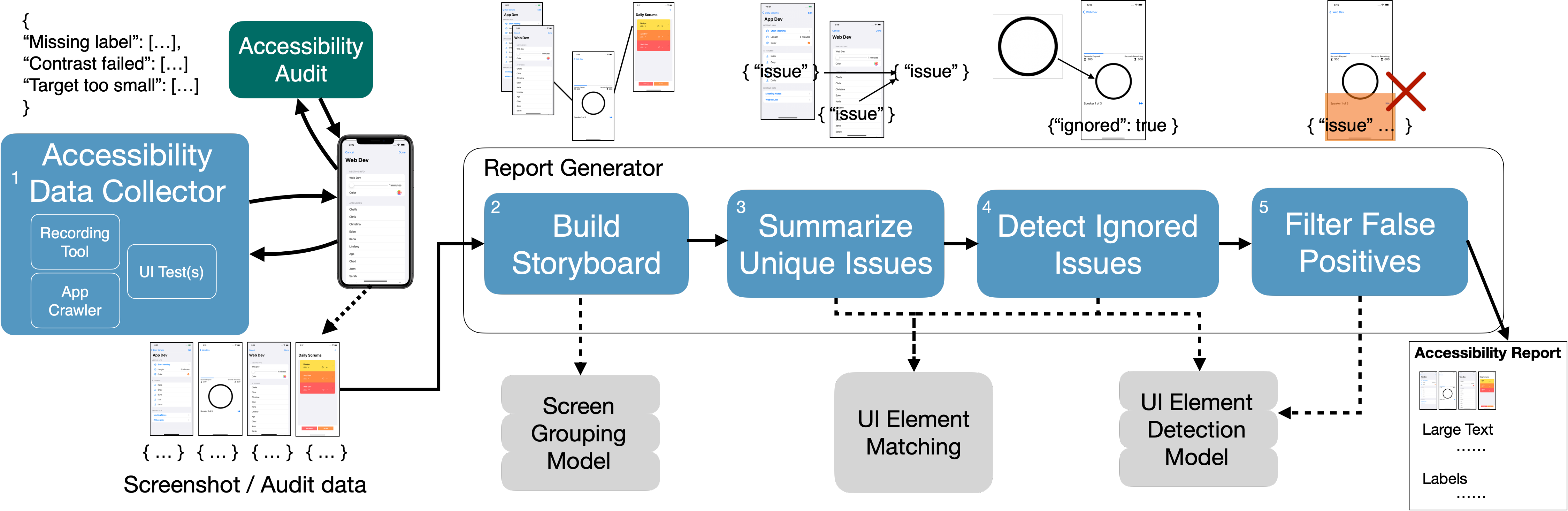}
    \caption{Our system for accessibility report generation. A data collector captures screenshots and accessibility audits, followed by a 4-stage process that produces a summarized accessibility report. We apply pixel-based machine learning models to de-duplicate and filter issues.}
    \Description{A flowchart diagram visualizing our system for accessibility report generation. The Accessibility Data Collector has multiple subcomponents including "Recording tool", "UI Test(s)", and "App Crawler". It runs on a mobile device and outputs a series of screenshots and accessibility reports. To the right, the chart visualizes the four stages of report generation including "Build Storyboard", "Summarize Unique Issues", "Detect Ignored Issues", and "Filter False Positives" with arrows going to each pixel-based model that they use. The output arrow shows a summarized report with screenshots and a list of issues detected.}
    \label{fig:system_overview}
\end{teaserfigure}
\maketitle

\section{Introduction}
Recent studies have found that a large number of both Android~\cite{ross2018examining} and iOS apps~\cite{zhang2021screen} are still missing basic accessibility. This lack of accessibility can result from misleading or missing labels that provide descriptions of UI elements to accessibility services, or result from UI elements being completely missing from their apps' accessibility meta-data and thus unavailable for interaction~\cite{zhang2021screen}. Why are developers not making their apps more accessible? For some, they may be unaware of accessibility requirements, and others may choose to deprioritize accessibility in favor of other app features~\cite{alshayban2020accessibility, vendome2019can}. 

Another cause may be a lack of efficient and effective accessibility testing tools. A variety of companies provide accessibility scanners, such as Accessibility Scanner for Android~\cite{googleAccessibilityScanner} and Accessibility Inspector on iOS~\cite{accessibilityInspector}, which must be manually activated on each screen of an app to dynamically test for a variety of accessibility issues. Unfortunately, it is laborious and time consuming for a developer to process and analyze a complete scan of their app. First, they must manually visit each screen in the app and collect a scan. Once they have completed scanning the whole app, developers must go through a lengthy process to examine the separate reports for each screen, identify true errors from false positives, and prioritize errors to fix. We conducted a formative study, and found that some developers analyze reports scan by scan, instead of in aggregate, creating duplicate work when there are overlaps between scans.  Most tools also have no memory from scan to scan, so each time they scan a screen, the developer must manually filter out false positives they have already identified as incorrect from previous scans.

Prior work has tried to address these challenges by providing accessibility app crawlers \cite{salehnamadi2022groundhog, eler2018automated, salehnamadi2023assistive} that randomly or through record \& replay approaches crawl an app to detect accessibility issues. There are two limitations to these approaches. First, they rely on accessible view hierarchies to drive the crawling itself which prior work has demonstrated to be often incomplete or unavailable for highly inaccessible apps \cite{zhang2021screen, li2022learning}, which are the kinds of apps we most aim to support with our system. Zhang et. al. \cite{zhang2021screen} found that 59\% of app screens had at least one UI element that could not be matched to an element in view hierarchy, and 94\% of apps had at least one such screen. Second, none of these works has yet studied how users interact with an interpret information from these accessibility reports, and what features are important in an accessibility report generation tool. 

In this paper, we introduce a pixel-based report generation system using an off-the shelf accessibility scanning tool which have have also instrumented to detect inaccessible elements with no corresponding match in the underlying view hierarchy. This can enable our work to report on a more diverse set of apps with incomplete or missing view hierarchies. Our accessibility report generation system uses a four step workflow that uses pixel-based machine learning models and heuristics to generate a high level summary of results and filter false positive issues (see Figure~\ref{fig:system_overview}).

To motivate our system, we interviewed 8 accessibility engineers and QA testers about their pain points in using current accessibility scanning tools. Through these interviews, we identified the following user needs for an accessibility report generation system: 
\begin{enumerate}
    \item Reduce the time required for developers to manually scan individual screens with accessibility auditing tools. 
    \item Provide developers with an overall app accessibility report. 
    \item Enable developers to reduce noise by ignoring false positive or previously addressed issues.
\end{enumerate}

Finally, we conducted a user study of our system where participants interacted with reports generated by a manual scanning tool and an app crawler. Participants were more satisified with their reports created by our system vs a baseline tool, and it helped them quickly prioritize important issues. Our study provides insights into the features that should be supported by accessibility report generation systems. 

 The contributions of this paper include:
\begin{itemize}
\item An improved screen similarity transformer model from [23] to 96.9\% accuracy (88.8\% f1-score) through a larger (6k apps) and more consistent dataset, which we leverage to generate application storyboards for reporting.
\item UI element matching heuristics achieving 97.3\% f1-score on 121k UI element correspondence annotations.
\item 3 key design goals for an accessibility report generation system, inspired by key limitations and inefficiencies with current accessibility scanning tools.
\item An accessibility report generation system instantiating these design goals by combining the screen grouping model with UI element detection~\cite{zhang2021screen} and matching to build an application storyboard of unique screens, de-duplicate issues, enable an ignore feature, and filter false positives.
\item A user study with 18 app developers and QA testers demonstrating that our report generation system can generate clean and accurate reports which can help them quickly prioritize and find common issues across an app. The study also reveals design insights for accessibility reporting interfaces and features needed to make them more effective in future systems.    
\end{itemize}

\section{Related Work}
Previous research systems have explored how to automatically collect, report, and repair accessibility issues. We also review work in methods for UI element and screen identification as our work improves upon these methods and enables use cases beyond accessibility report generation. 

\subsection{Automated Tools for Accessibility Analysis}
Several tools exist to check accessibility properties of apps~\cite{silva2018survey}, and they can generally be categorized as development-, run-, or test-time. 
Development-time tools, such as Android Lint~\cite{googleAndroidLint}, use static analysis techniques to examine code and declarative user interface descriptions for potential issues. These tools do not have access to user interface elements that may be created programmatically or data that is downloaded at run-time. 

Run-time tools~\cite{googleAccessibilityScanner,googleEspresso,robolectric2021,accessibilityInspector} examine the running user interface in the same way as a user would experience the app and can find issues that development-time tools would miss~\cite{eler2018automated}, but may be limited by the capabilities of the automated system exploring the app.
Some common run-time tools, like Accessibility Inspector for iOS~\cite{accessibilityInspector} and Accessibility Scanner for Android~\cite{googleAccessibilityScanner}, require developers to visit and scan each screen of an app. The results are separate reports for each screen, which must be manually analyzed with no option to generate or view an overall summary of issues as in our work. 

Run-time analysis can also use automated crawling on a running app to find accessibility issues. To our knowledge, all past systems work on the Android platform and rely on a view hierarchy and accessibility meta-data to understand the app contents. While some common Android components are accessible by default (e.g., View), this is often not the case for custom widgets.
MATE~\cite{eler2018automated} uses dynamic, random exploration of Android apps, relying on UIAutomator, and detects accessibility issues on encountered screen states. Both Xbot~\cite{chen2021empirical} and Alshayban et. al.~\cite{alshayban2020accessibility} crawl apps to collect accessibility issues but either rely on app instrumentation or static analysis of source code to extract intents, which they note will only work for a limited number of Android apps~\cite{chen2021empirical}. All three~\cite{eler2018automated, chen2021empirical, alshayban2020accessibility} appear to rely on Android Activity and heuristics to determine screen states, and primarily focus on generating issue counts. More recently, the accessibility app crawler Groundhog~\cite{salehnamadi2022groundhog} alternatively crawls apps through accessibility services to detect additional classes of issues (i.e., locatability, actionability), producing a report of issues, and a video to visualize navigational failures (i.e., talkback). However, the paper does not provide details on the interface for the output report or study how users interpret issues from it. Our system uses an existing accessibility scanning tool within an app crawler, thus reports on more and different classes of issues than Groundhog. By leveraging a pixel-based app crawler, similar to ~\cite{wu2023neverending}, our system can navigate to areas of the UI that would be inaccessible through accessibility services, which Groundhog relies on. Prior studies~\cite{zhang2021screen} have demonstrated a large amount of apps still have many UI elements and screens that are not exposed to the accessibility hierarchy. In contrast to prior works, we also study how to report and summarize detected issues to make them interpretable and actionable through a user study with app developers and QA testers.

Test-time tools~\cite{googleEspresso, robolectric2021, earlGrey2022, salehnamadi2021latte} are integrated into functional or user interface testing processes, and collect data when tests are run. These tools also collect data from the running user interface to find issues that development-time tools would miss, but may be limited by the completeness and coverage of the tests. Unfortunately, past work has shown that mobile apps are often tested in an ad-hoc manner~\cite{cruz2019attention} or not at all~\cite{kochhar2015understanding}. Latte~\cite{salehnamadi2021latte} eases the process of accessibility test creation by working with test cases written for functional UI correctness, which are easier to author than UI integration tests. Latte tests for accessibility by replaying test cases through available accessibility services, such as SwitchAccess or TalkBack on Android. While Latte can detect more accessibility problems than prior work and works with a larger variety of test cases, it remains only as effective as the coverage of the test cases across the entire app. 

In summary, prior approaches have detected accessibility errors through different ways of exploring an app: manual capture, integration with existing UI tests, or automated app crawlers. We instead focus on assembling the results of accessibility error reports, agnostic to the collection method, into a single app report with an overall summary. We also present new methods to summarize unique issues, filter false positives, and enable developers to ignore issues in future reports, which can help reduce noise when adopting such systems to monitor for accessibility regressions over time.

\subsection{Improving Accessibility through ML}
Our approach relies on ML-based screen and UI understanding techniques to summarize accessibility issues, detect false positives, and produce an actionable report. Machine learning techniques have recently been explored to improve app accessibility by detecting UI elements and exposing them to screen readers~\cite{zhang2021screen}, generating labels for app icons to expose to accessibility services~\cite{chen2020unblind, zhang2021screen, chen2022icons, mehralian2021data}, repairing size-based accessibility issues~\cite{alotaibi2021automated}, and detecting visual display issues~\cite{liu2020owl}. These approaches demonstrate that machine learning can generically detect and even repair accessibility problems. In our work, we apply previous work on UI element detection~\cite{zhang2021screen} to detect and filter issues in our generated accessibility reports. While our system focuses on reporting over repairing, it could be used in combination with prior work to detect issues and alert developers to fix problems before releasing their app, possibly reducing the need for dynamic repair. 

\subsection{UI \& Screen Identification}
Our work presents a new model for screen identification and heuristics to identify UI elements across different instances of a UI screen within an app. This work contributes to a body of work in screen and UI understanding in helping developers and designers understand and explore the structure of their own and similar apps. 

Across apps, some work targets search in large UI datasets. Gallery-DC~\cite{chen2019gallery} presents a searchable gallery of UI components using a deep-learning based object detection approach.  Liu et. al.~\cite{liu2018semantics} present a model to learn screen level embeddings from semantic UI element annotations, which can be used to search UI components and screens in a large dataset. At the screen level, Rico~\cite{deka2017rico} presents an autoencoder to search for similar screens in a large UI dataset. VINS~\cite{bunian2021vins} applies object detection image-based retrieval to help designers find examples. Screen Parsing~\cite{wu2021screenparsing} detects UI elements with object detection, and infers layout structure using a transition-based parser to enable screen similarity search.  Swire~\cite{huang2019swire} applies a deep neural network to sketch-based UI image retrieval. While some ML aspects of these works may aid in our screen grouping problem, they apply screen similarity detection across similar screen types in different apps, while we aim to group screen types within an app. Additionally, some of this work~\cite{deka2017rico, bunian2021vins, wu2021screenparsing} does not appear to incorporate visual information into the similarity problem, which we believe can provide important cues for same screen detection.

For same screen detection within an app, prior work has proposed heuristics-, modeling-, and hashing-based approaches. Earlier works applied a perceptual hash~\cite{gianazza2014puppetdroid} to detect the same screens within an app, but other later work showed that hashing techniques have high precision but very low recall~\cite{feiz2022understanding}.  In accessibility report generation, this type of performance could result in much noisier and less usable reports. Zhang et. al.~\cite{zhang2018robust} present screen and UI element equivalency heuristics based on identifiers and structures in Android view hierarchies. App crawling, a key use case for same screen detection, relies on similar heuristics based on view hierarchy structures~\cite{chen2018fragdroid, jiang2018makes, li2019humanoid, li2017droidbot} which prevents them from being generalizable to other platforms. ~\cite{feiz2022understanding} presents a machine learning approach to same screen detection within an app. We build on this work, but use a modified definition of the `same screen' problem.

\section{Background \& User Interviews} \label{section:formative_interviews}
This project began as a collaboration with our research team, the accessibility engineering team from a large technology organization and a product manager in charge of app accessibility for a different team in the same organization. From these stakeholders, we initially learned about the challenges of collecting and assembling accessibility reports for a full app. To understand more about these challenges from a larger group, we interviewed 8 accessibility-focused developers, testers, and managers from 5 diverse product and research-focused teams at the same organization. Through a small set of structured questions, we discussed their experiences testing app accessibility in 30 minute exploratory interviews.

First, we asked participants to describe which parts of testing for app accessibility they found challenging and what they liked and disliked about accessibility scanning tools. The primary tool our participants had used was Accessibility Inspector~\cite{accessibilityInspector}, however, participants also mentioned using the Evinced scanner~\cite{evinced}, Lighthouse~\cite{googleLighthouse}, and Android tools (e.g., ~\cite{googleAccessibilityScanner}). While the participants sometimes wrote accessibility tests and used automated scanners, they reported primarily manually testing their apps. 

\textbf{Current tools provide no results overview:}
Our participants mentioned that since scanning tools provide results per screen, they can't easily see an overview of results -- \textit{``I can't really get an overview of an app's accessibility just from that tool''} or \textit{``view the issues of a particular type across the app''}. Some participants said it can be hard to give feedback on app accessibility to teams that may not understand accessibility well or know what to test. Such teams might benefit from feedback on issue patterns across the app (e.g., missing Large Text support), which for some participants to compile may require \textit{``toggling on that feature and navigating through every single screen of the app myself to get an idea of whether this app does or does not support those [accessibility features]''}

\textbf{Current tools are too noisy:}
Participants mentioned that results of current tools \textit{``can be quite noisy at times and so we end up with a lot of false positives"}. For someone less familiar with accessibility features, \textit{``they have a really hard time understanding what's signal and what's noise from the report''}. Participants also mentioned many issues detected by these tools are lower priority to fix --~\textit{``false positives are confusing. There's definitely a difference between elements that can't be visited in any way, and are totally inaccessible, compared to some of the smaller nit picks that get presented."}

\textbf{Manual scanning introduces inefficiencies:}
Participants also recounted the manual effort and time to use accessibility scanning tools \textit{``it will take me a couple hours just to get through a couple screens, like a few screens usually''}. The amount of effort involved often leads them to scan their apps infrequently. When multiple developers or teams contribute to the same app, manually scanning after each change is not possible, so our participants conduct scans infrequently. Accessibility regressions can be created and persist for quite some time. It can also be infeasible to run these scans across the multitude of combinations of devices and accessibility settings they would like to test.

.

\subsection{Design Goals}
From these formative interviews, we formulated the following design goals for accessibility report generation system. 
\begin{itemize}
    \item \textit{D1}: Reduce the need for developers to manually scan individual screens with accessibility auditing tools. 
    \item \textit{D2}: Provide developers with an overall app accessibility report.
    \item \textit{D3}: Enable developers to reduce noise by ignoring false positives or previously addressed issues.
\end{itemize}

For the first goal, we adopt an app crawler that was introduced in prior work \cite{wu2023neverending} that we modified to audit each screen using the Accessibility Inspector \cite{accessibilityInspector} and produce an output HTML report in a live webpage. We also provide a manual tool for accessibility auditing which generates a multi-screen and developed features to ignore false positives \& previously addressed issues which are available in the web report.

\begin{figure*}
\includegraphics[width=\linewidth]{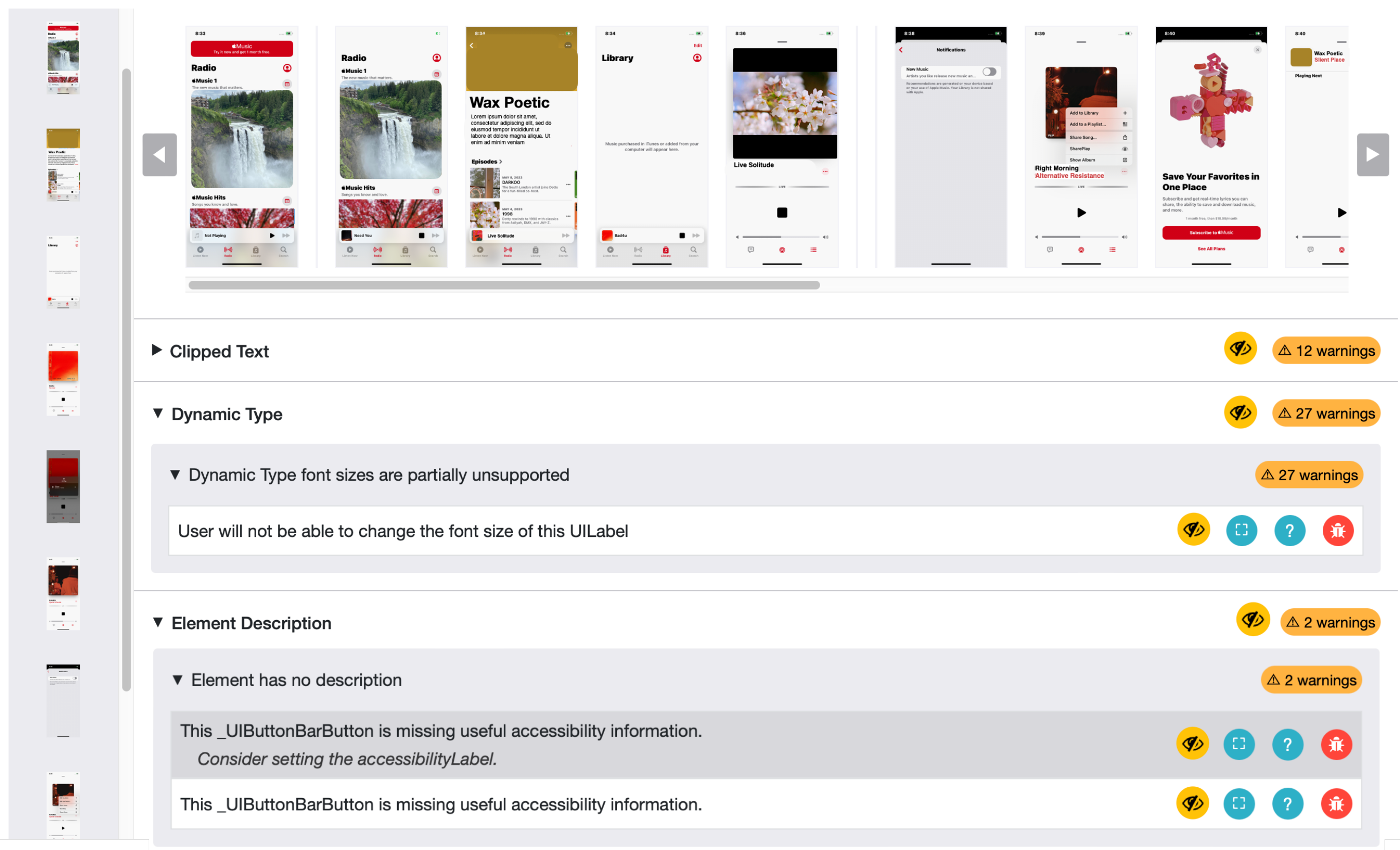}
  \caption{The prototype interactive HTML report interface generated by our report generation system.}
  \Description{A screenshot of our interactive report interface. On the left, it contains vertical tabs including a Summary tab and a tab for each screen group. The summary tab is currently open, and displays a row of various screenshots from the app across the top with a summary of issues listed below. The interface categorizes issues by their issue type, like Hit Region and Large Text, and displays a counter next to each header to see the number of issues in that category. Underneath each category, the interface lists subcategories of types of issues (e.g., for Hit Region - "Hit area is too small"), and ignore buttons next to each issue type. Each issue in the subcategory has a button next to it to file a bug, see more info about how to fix the issue, and an ignore button.}
  \label{fig:report_ui}
\end{figure*}

\section{Accessibility Report Generation}
\label{section:system}
Figure~\ref{fig:system_overview} illustrates our approach to generate accessibility reports. First, a data collector, such as manual capturing tool, an app crawler, or a test case-based recorder, captures screenshots and accessibility data (Figure~\ref{fig:system_overview}.1). Next, a \textit{report generator} module generates a summarized report grouped by screen types detected by a screen grouping model (Figure~\ref{fig:system_overview}.2-5). The report generator then uses UI element de-duplication heuristics to de-duplicate current issues and filter previously ignored issues. Lastly, the report generator uses a UI element detection model~\cite{zhang2021screen} to filter false positives and produce a summarized report. 

We implemented our prototype as a Flask-based web server that controls data collection via a proprietary device cloud, generates reports, and hosts reports for later viewing by users. We currently only support generating accessibility reports for iOS-based apps. 

\subsection{Accessibility Data Collection}
The first step of report generation is capturing accessibility data to report (Figure~\ref{fig:system_overview}.1). This is aimed towards design goals \textit{D1} and \textit{D2}. In our prototype, we offer two options for data collection: a manual auditing tool and a random app crawler using the architecture adapted from ~\cite{wu2023neverending}.

The random app crawler runs on a remote cloud device or a locally attached iOS device. The random app crawler detects clickable UI elements on each screen using UI element detection~\cite{zhang2021screen}, and interacts with them to explore the app. It uses our screen grouping model to find new screens to attempt to maximize coverage. On each screen, it captures an accessibility audit and screenshot and may capture a screen in various states. This is a quicker option to generate a report, but provides no guarantees of obtaining complete coverage of the app. 

Users interact with the manual auditing tool via desktop MacOS interface. The interface connects to a locally attached device or simulator and provides a button ``Run Audit'' to capture an accessibility audit and screenshot. While this does not directly meet design goal \textit{D1}, the user cannot examine reports as they are generated, and may be less likely to get distracted by the results until they have finished capturing. Once finished, the system generates a summary report for all screens captured by the user (\textit{D2}). Using this data collection requires manual effort, but gives the user complete control over what screens are captured. 

Both of these methods collect accessibility audits and screenshots. We currently use Xcode's Accessibility Inspector feature via a command line tool on each device, which produces a JSON report listing all detected issues with their associated bounding box on the screen. The Accessibility Inspector supports 29 accessibility checks, categorized by \textit{Element Description}, \textit{Contrast}, \textit{Hit Region}, \textit{Element Detection}, \textit{Clipped Text}, \textit{Traits}, and \textit{Large Text}. In the future, it should be possible to add other audit tools to our capture process, provided they can run on a live device and produce JSON output.

\subsection{Building a Storyboard}
\label{section:screen_grouping}
Data collectors may often capture multiple instances of screens that appear slightly differently, perhaps because they are scrolled or contain some dynamic content, but include duplicates of the same accessibility issues. If we reported each screen instance individually, the overall report would be noisy and contain many duplicate reported issues, which was a major concern from our formative interviews. To mitigate this, the report generator builds an app storyboard (Figure~\ref{fig:system_overview}.2) using a screen grouping model which groups together the results from different instances of the same screen from their screenshots alone. We opted to use a pixel-based model only rather than relying on view hierarchies as in prior work~\cite{salehnamadi2022groundhog} which can enable our system to work on apps without a view hierarchy available. In this work,  we adopt the term “storyboard” from UI builders (e.g., Xcode Storyboards) which use “storyboard” to describe a visualization of relationships between views in an app, rather than the definition of "storyboard" to convey a user story in UX design. 

\begin{figure*}[!htb]
    \centering
    \includegraphics[width=\textwidth]{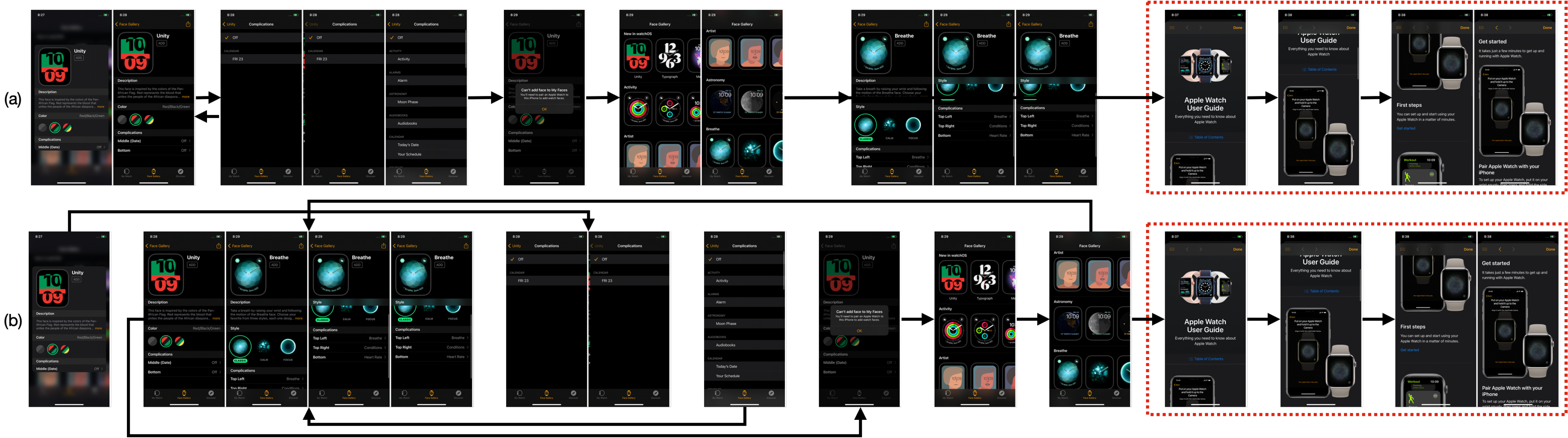}
    \caption{Examples of partial app storyboards generated by our screen grouping models: (a) Similarity Transformer, (b) Bi-Encoder. The red box marks a missed opportunity for grouping in (a) where the screen has been scrolled. The full app storyboards for this crawl are included in the appendix.}
    \Description{Two example app storyboards are shown. The first row shows a similarity transformer generated storyboard of a Watch app with several groups having small variations among the same screen instances (e.g., scrolling, different content) while still being grouped correctly. Some screens in the far right of the storyboard are different instances of the same screen scrolled down and have not been grouped. A red box is around them to highlight the error. The storyboard also shows arrows between various screen groups based on where they were encountered during the crawl. The bottom row shows a storyboard of the same app generated by the bi-encoder model which has more groups. While some screens are grouped together well, several screens are individually grouped.}
    \label{fig:storyboard_examples}
\end{figure*}

\subsubsection{Screen Grouping Model}
Screen grouping is a key technology to generate an app storyboard, summarize accessibility issues, and detect ignored issues (Figure~\ref{fig:system_overview}2-4).  If the grouping model performs poorly, then the report will likely contain more duplicates that users need to sort through and decrease their confidence in the usefulness of the report. To help mitigate this risk, we developed an improved  screen grouping model compared to previous work \cite{feiz2022understanding}.  We combine this model with the UI element matching heuristics described in Section ~\ref{section:ui_element_fingerprinting} to generate a summarized report for each screen group. 

For storyboard generation, we first trained a similarity transformer model building on previous work. We contacted \citet{feiz2022understanding} and they agreed to share their dataset and labels for training. The input to this model is two screens, $s_{1}$ and $s_{2}$, and the output is a binary prediction of \textit{same screen} or \textit{different screen}. Despite various experiments with model parameters, we were unable to improve this model's performance beyond 75.8\% F1-score when trained and evaluated on its original dataset. Note that this F1 score is lower than previously reported~\cite{feiz2022understanding}, we believe due to different choices in training, validation and test set splits. We discovered the annotation process in the original work had a low agreement rate. To address this, the authors dropped many app crawls from their final dataset. We subsequently focused our efforts on improving the annotation process, both to increase annotation agreement and to make more data available for training and evaluation. 

We examined 1000 failure cases from the similarity transformer model by computing screen comparison predictions for each pair of screens in a validation set. We categorized each pair into cases where either the model produced incorrect predictions, or where the annotators grouped them incorrectly. Model prediction errors occurred primarily when two screens were the same but were scrolled or had structural differences (e.g., keyboard open / closed, search with and without data) and were predicted as different, or the opposite for two different screens that were structurally and visually similar. The model was often confused when a dialog box or other overlaid foreground element was present. Annotator disagreement occurred when screens had different tabs or page controls active, and when the same screen displayed different data (e.g., the same profile screen with a different person shown). 

\subsubsection{Screen Grouping Data Collection}
To improve the quality of this dataset for our report generation use case, we collected a new labeled dataset of 750,000 grouped screens from 6700 free apps using a similar data collection process to~\citet{feiz2022understanding}. To collect the 750k app screens, crowd workers manually explored apps through a remote device in a web interface with the instruction to find as many unique screens as possible within a 10 minute limit. While workers were crawling, the system captured screens every second provided the screen changed.

A different set of 15 crowd workers grouped the screenshots of each app into \textit{same screen} groups using a card sorting style interface similar to that used by~\citet{feiz2022understanding}. Crowd workers could drag and drop screens into various groups to combine them into same screen clusters. Rather than having the crowd workers start with the screens completely ungrouped, we generated initial groupings using our initial trained screen similarity transformer with 75.8\% F1 -score and the \textit{Build Storyboard} module from our report generation. Thus, crowd workers only needed to fix the model's mistakes rather than starting from scratch. We also added a special ``trash'' group to the annotation interface to discard invalid screens (e.g., home screen, loading screens, landscape orientation). 

In our annotation guidelines, we defined \text{same screen} as two screens used for the same purpose, to accomplish the same task, or to view the same type or category of information. From our observations of the model errors, we defined a list of possible variations a screen can undergo to be considered the same screen including: 
\begin{itemize}
\item Same screen with different data
\item Partially scrolled down 
\item Sections expanded or collapsed 
\item Keyboards open or closed
\item Non-modal application dialog open or closed
\item Same modal menus or dialogs on top of different content. 
\end{itemize}

In contrast to~\citet{feiz2022understanding}, we define same screens in terms of the topmost layer of interactive content rather than the background layer. For example, if two screens have the same modal dialog open over different backgrounds, we consider them the same screen since the screen behind is non-interactive. We define left sliding ``drawer'' menus similarly. Grouping screens this way is more applicable to the report generation task as nearly all accessibility audits focus on the topmost, non-occluded layer (e.g., contrast checks).  After training annotation workers with these guidelines, they produced screen groupings for each of the 6700 apps. The labeled data contains 70,882 groups, with a mean of 10.8 groups per app (Med: 10, Std: 7.3) and 3.3 screenshots per group (Med: 2, Std: 6.7). Around 700 apps of our labeled data consisted of entirely ``trash'' screens (e.g., landscape, loading), and we did not include these in our final dataset. 

A separate group  of 5 expert QA annotators reviewed 10\% of the annotations for accuracy. If any batch of the annotations did not exceed 98\% acccuracy as reviewed by QA, the batch was sent to be re-annotated until exceeding the 98\% accuracy threshold. 

\subsubsection{Screen Grouping Model Training}
To understand the characteristics of our model and performance implications of our new annotations, we trained two additional similarity transformer models to compare with the initial transformer trained above, including a model using the architecture from~\cite{feiz2022understanding} and a modified version that produces an embedding for each screen.  To ensure the results were comparable, we split the data by app into training, validation and test sets.  We split by app to ensure that screens from the same app appear in only one set.  We use identical splits for both the original data annotated by~\citet{feiz2022understanding} and a 1k subset of our newly annotated data, which we term FD and 1k respectively. We also created a 6k dataset containing all crawls, where we added all crawls not in the 1k dataset to the 6k training set. Thus the same set of apps we include in the 
validation and test sets for all datasets. We trained the transformer models on the 1k and 6k datasets. The training input to each transformer are pairs of "same" and "different" screens generated from the groups of each crawl. The full data is significantly unbalanced, containing 8.2 million pairs of "different" screens and 3.3 million pairs of "same" screens. We present the details of the evaluation results in Section~\ref{section:screen_grouping_eval} 

We trained the first transformer model as a cross-encoder which predicts the similarity label for a pair of input screens by minimizing the binary cross-entropy loss on the predicted similarity label, similar to~\cite{feiz2022understanding}, and apply a similar masked prediction objective. For the embedding model, we train a bi-encoder that uses a transformer network to generate an embedding for each screen by encoding and pooling the output of pre-trained object detection model for the screenshot. During training the model learns to minimize the distance between the embeddings of similar screens while maximizing the distance for different screens. We also apply the masked prediction objective to the bi-encoder model which we found to improve performance in our experience.

\subsubsection{Storyboard Generation}
To build a storyboard using this model (Figure~\ref{fig:system_overview}.2), the report generator processes each screen consecutively. When it adds a new screen to the set, it compares the new screen to a representative screen from each group found so far. Each model prediction produces a float value (positive if \textit{same screen} is predicted) and (negative if \textit{different screen} is predicted). The system assigns the screen to the group with the highest positive score, or to a new group if the model doesn't predict any positive scores. To create transition edges, the report generator maintains a $current\_group$ variable with the last encountered group. When the report generator adds a new screen to an existing group $e$, it adds an edge between $current\_group$ and $e$, and sets the $current\_group$ to $e$. When the report generator adds a new screen to a new group $n$, it adds an edge between $current\_group$ and $n$ and sets $current\_group$ to $n$. Our interface (Figure~\ref{fig:report_ui}) displays this storyboard on a separate tab in the report (not pictured but similar to Figure~\ref{fig:storyboard_examples}). 

\subsection{Issue De-duplication}
\label{section:ui_element_fingerprinting}
A key requirement of report generation motivated from our interviews is to avoid noise (design goal \textit{D3}). We learned in formative work that accessibility scanning tools can often produce noise on a single screen (see Section~\ref{section:formative_interviews}), and a report generator should reduce this noise when summarizing results across multiple screens. After the \textit{Build an App Storyboard} step, our report generator (Figure~\ref{fig:system_overview}.3-5) applies UI element matching heuristics to identify multiple instances of the same element across multiple screens, remove any duplicate accessibility issues created by inspecting the same element multiple times, and summarize results across an app and on each screen. The system applies these same heuristics to ignore issues, marked by users in the report interface (Figure~\ref{fig:report_ui}) and re-identify them on future runs. To enable this, we developed robust heuristics to \textit{find the same UI element across two different instances of the same screen within an app}.

The input to UI element de-duplication is a pair of screens which our screen grouping model has detected as the same but may have some variations (e.g., scrolled down, keyboard open). The pair of screens consists of a \textit{template} screen $T_{s}$ and UI element $T_{ui}$, and a new screen $N_{s}$. The goal is to find the best matching UI element $N_{ui}$ within the new screen $N_{s}$ for a template UI element $T_{ui}$.

\begin{figure}[t]
    \centering
    \includegraphics[width=0.75\linewidth]{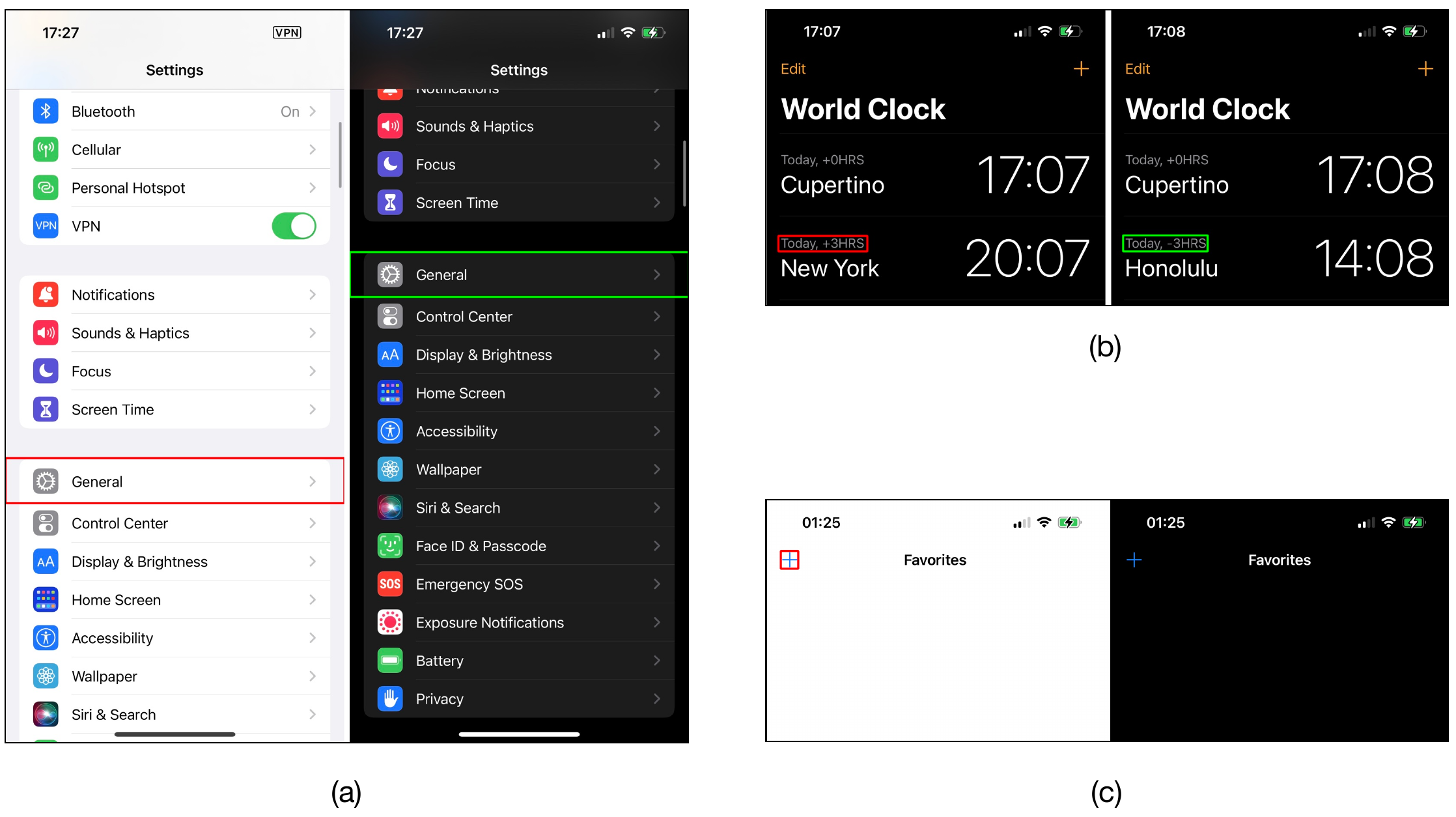}
    \caption{Examples of True Positive (a),  False Positive (b), and False Negative (c) of UI element de-duplication heuristics. The red box indicates target UI on original screen, while the green box indicates the matched UI on the new screen.}
    \Description{Three examples of the results of UI element de-duplication heuristics. Each example shows a screen pair with a red box highlighting the target UI element on the left screen and a green box highlighting the matched UI element. The first example (a) shows a True Positive on a Settings screen where the right screen has been slightly scrolled up and the target UI element is detected correctly. The second example (b) shows a False Positive where the text "Today +3HRS" above two different cities has been matched on a World Clock screen. The third example (c), shows a False Negative where a "plus" icon in the top left corner on a white background has not been matched in the new screen which has a black background.}
    \label{fig:ui_element_matching_examples}
\end{figure}

\begin{figure}
    \centering
    \includegraphics[width=0.75\linewidth]{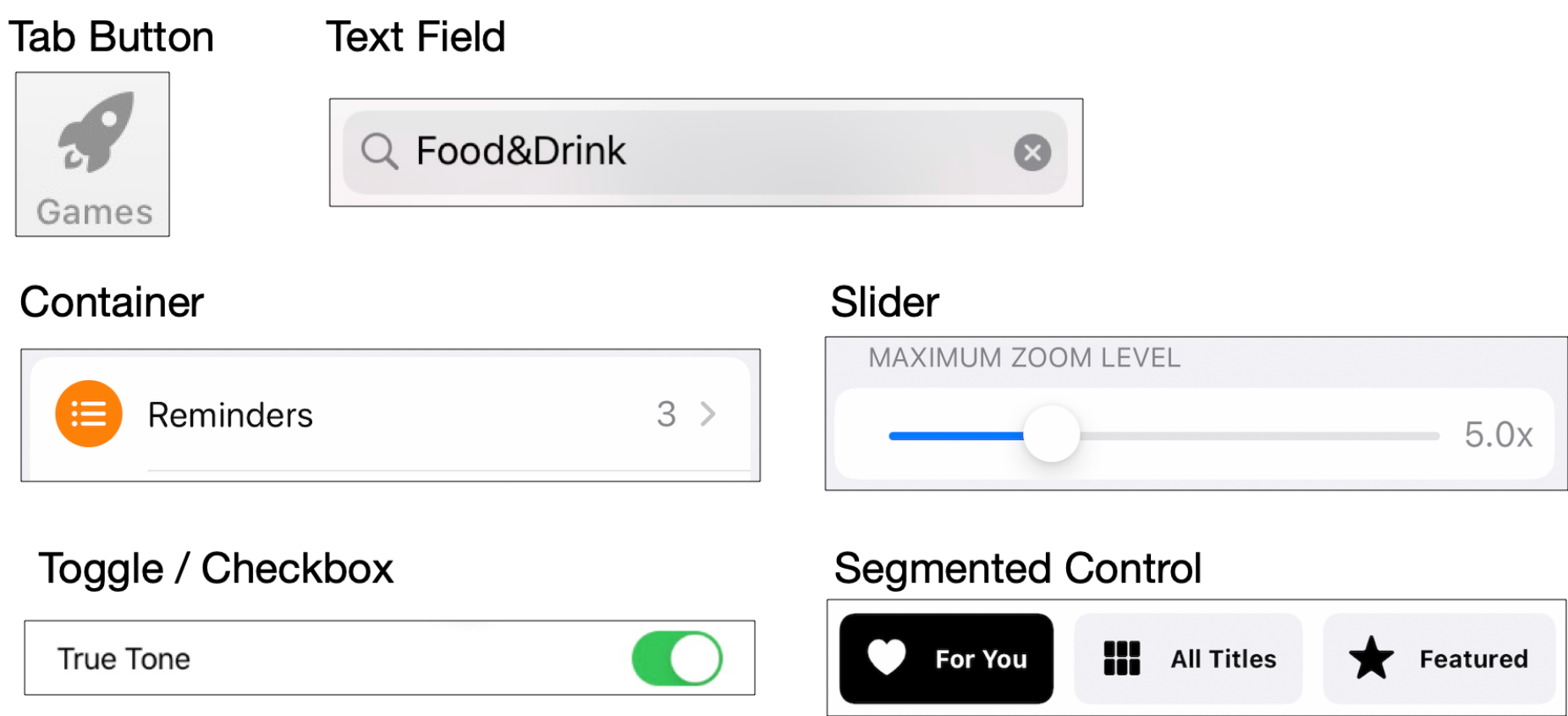}
    \caption{Groupings used by UI element fingerprinting heuristics to find UI element detections in a new screen.}
    \Description{Examples of UI element groupings used by UI element de-duplication heuristics to find a template UI element. They include TabButton, TextField, Container, Slider, Toggle/Checkbox and SegmentedControl examples with each high level UI element (e.g., TabButton) grouped with its associated contents (i.e., Icon and text label)}
    \label{fig:ui_fingerprinting_groupings}
\end{figure}

\subsubsection{Pre-processing of template screen and UI element}
\label{section:preprocessing}
UI element matching first gathers information to compare $T_{ui}$ to elements in the new screen by pre-processing the template screen $T_{s}$ and template UI element $T_{ui}$ using the following steps: 
\begin{enumerate}
    \item Detect all UI elements on $T_{s}$ with a UI detection model.
    \item Locate $T_{ui}$ in UI element detections.
    \item Create groups for UI element detections.
    \item Save UI element detections, screenshot of $T_{s}$, groups, and $T_{ui}$
\end{enumerate}

In Step 3, we adapt the heuristics introduced by ~\cite{zhang2021screen} to group UI elements detected by the UI detection model as follows (see Figure~\ref{fig:ui_fingerprinting_groupings} for examples):

\textbf{Tab Button}: A tab button group often contains a icon and text, and sometimes contains only a icon.

\textbf{Toggle or Checkbox}: A toggle (or checkbox) group contains that element and its text description, which is often the closest text on the same row.

\textbf{Segmented Control}: A segmented control group contains the border and the text of a segmented control.

\textbf{Text Field}: A text field group contains the border of the text field and UI detections inside it.

\textbf{Slider}: A slider group contains the slider and text on the same row and the closest text above.

\textbf{Container}: A container group contains the border of the container and UI element detections inside it.

\subsubsection{Finding the best matching UI element}
Our system applies the following heuristics to find the best match for a template UI element $t_{ui}$ in a new screen using a \textit{findBestMatch} method: 

\begin{enumerate}
    \item Get all UI detections on new screen $N_{s}$. 
    \item Create groupings of UI element detections on $N_{s}$ (as described in section~\ref{section:preprocessing}). 
    \item Compare each UI detection with $T_{ui}$ using \textit{matching heuristics} and get their similarity. 
    \item Pick the UI element with the highest similarity score, if at least one match candidate is found. 
\end{enumerate}

Our system applies the following matching heuristics based on the detected UI type of $T_{ui}$: 

\textbf{Text}: Pre-process the text to make it lowercase and keep only alphanumeric characters and spaces. The similarity score is the score of text fuzzy matching \cite{rapidFuzz2022}.

\textbf{Icon and Picture}: Search the area around the Icon or Picture detection using image template matching ~\cite{opencv} (template = the cropped pixels of $T_{ui}$). Our method creates the template in multiple scales\footnote{S = $\frac{T_s width}{N_s width}$; Scales = [0.91*S, 0.94*S, 0.97*S, 1.0*S, 1.03*S, 1.06*S, 1.09*S]} so when two screenshots are the same size, our method can still find the match in different scales. The similarity score is the max value of template matching among all scales.

\textbf{Tab Button}: If the template Tab Button contains only Icon, run the Icon matching method to compare the Icon. When it contains both Icon and Text, run the Text matching method. 

\textbf{Toggle, Checkbox, Segmented Control, Slider and Text Field}: Run the Text matching above on its grouped Text.

\textbf{Page Control and Dialog}: Normally, there is at most one Page Control or Dialog on a screen. When there are multiple of the same type, our method compares the distance (normalized with screen width) between $T_{ui}$ and $N_{ui}$. The similarity score is $1 - distance$.

\textbf{Container}: If a Container only contains Icons, run the Icon matching above. Otherwise, run the Text matching above on each Text inside the Container (in reading order).

To pick the best match, our system looks for the UI element with the highest similarity score. If no UI elements pass a threshold\footnote{Text: 90\%, Icon: 80\%, Picture: 50\%}, then there is no match. We determined these thresholds empirically using a large dataset of same screen pairs and annotated UI elements that we describe further in section~\ref{section:data_collection_annotation}. When a Text or Icon is in a grouping, our method first determines if the grouping is a match, and then prioritizes candidates in the same grouping.

\subsection{Building the Final Report}
Finally the report generator completes the report by detecting and hiding previously ignored issues and any false positive issues that the user has marked in the UI (Figure~\ref{fig:report_ui}). This is accomplished in two steps using the screen matching and UI element matching methods described in previous sections. Both detecting ignored issues and filtering false positives target design goal \textit{D3} of reducing noise in the report by ignoring issues and hiding false positives. 

First, the report generator retrieves the ignored issues, elements, and screens from a database. A report user would have previously saved these using the interface pictured in Figure~\ref{fig:report_ui} by clicking the eye slash button. For each ignored issue on a screen, the report generator finds the matching screen in the summarized report with the screen grouping model. On the matching screen, the report generator uses the \textit{findBestMatch} method to find the best matching UI element. If this UI element has any matched issues of the same type as the ignored issue, the report generator marks them as ``ignored'' and moves them to a collapsed section (Figure~\ref{fig:system_overview}.4).

If the user marked any screen from a previous report as ignored, the report generator compares each screen in the report to the stored \textit{ignored} screen using the screen grouping model, and moves any screen detected as a \textit{same screen} to an ignored section.

Second, the report generator filters false positives using a basic heuristic (Figure~\ref{fig:system_overview}.5). Any issue produced by the auditing tool with no visible matched UI element detection is assumed to be a false positive and the report generator hides it in the output report.  

The output of the report generator is a self-contained report in a JSON file, containing a summary of unique issues detected across each screen group of the app. The report generator categorizes issues by accessibility issue category (e.g., Element Description, Dynamic Type) and subcategory (e.g., Element has no description, Dyanmic Type partially unsupported).

\subsection{Report User Interface}
In our prototype, we convert the JSON file produced by the report generator into a prototype interactive HTML report summary, shown in Figure~\ref{fig:report_ui}. In the future, it could be rendered as a static report or further processed in a continuous integration pipeline.

The interface displays an overall summary tab to explore all issues discovered across the app and tabs to visualize results within each screen group. The interface categorizes issues by type (e.g., Element Description) and provides an overall count for each category. A person examining the report can click categories or issue headers or rows to display all screens impacted by the activated issue. The report interface additionally highlights each impacted UI element on each screen. To view the results for a specific screen group, the report user can click through each screen tab along the side,  which we currently visualize with a small thumbnail image of the screen. When clicking on each tab, the report presents a similar summary interface for each screen group. 

The interface provides several options to facilitate triage and reduce noise in future reports (design goal \textit{D3}). First, the report user can directly file a bug with a bug tracking system for each detected issue (through the Bug button). The user can also click a button to ignore a specific issue, issue type, category, or screen in a future report, which are saved into an ignore issue database. The interface enables removing these ignores at any time through a separate section. For any screen or issue ignored, the interface hides these in future reports for this app using screen and UI element de-duplication as previously described. Next to each issue row, the interface also provides a question mark button to view more information on suggested fixes for the issue.

\section{Technical Evaluation}
\label{section:technical_evaluation}
We conducted technical evaluations of two parts of our system, including the screen grouping model and the UI element de-duplication heuristics. For the screen grouping model, we report the overall results on the test dataset, shown in Table~\ref{table:screen_sim_model_results}, and for UI element de-duplication, we collected and evaluated the heuristics on a large evaluation dataset. 

\subsection{Screen Grouping Model}
\label{section:screen_grouping_eval}
Overall, we can see that the transformer trained on 6k outperforms the other models, though the transformer trained on 1k performs surprisingly similarly. The transformers trained with new annotations also perform noticeably better than those trained on the old annotations. To more deeply explore the impact of the different annotations, we also evaluated the initial transformer trained on FD with the test set from 1k. Interestingly, the model trained on FD performs better on the 1k test set than the FD test set, which may indicate it was able to learn some concepts through annotation noise in FD that were more applicable in 1k.

While this model can predict same screens with high accuracy and generates reasonable looking app storyboards (see Figure~\ref{fig:storyboard_examples}.a), it can be computationally costly ($O(n^2)$) as each screen in a set of captured screenshots is added. 
Reports cannot be generated interactively when this model is combined with the other report generation processes (e.g., summarizing issues), because data collectors and report generation may take a few hours to complete. As our ultimate goal is for such reports to be generated on demand, ideally within a few minutes, also trained an embedding version of the transformer model for screen similarity. 

Embedding based approaches can be more efficient as embeddings can be computed a priori and systems can calculate the distance between embeddings to determine similarity versus conducting a pairwise inference with all known screen groups. To investigate the impact of this approach on storyboard generation, we trained a screen similarity embedding model based on the same architecture as~\cite{feiz2022understanding} which produces a fixed-size embedding for each screen. We trained this model on the same splits we trained the similarity transformer on for the 6k dataset,

Using this model, storyboard generation can compute the Euclidean distance between the new screen embedding and a the mean of the embeddings for a group, as compared to the similarity transformer which requires a model prediction. Table~\ref{table:screen_sim_model_results} summarizes the performance results, with a distance threshold of 0.2 to determine if two screens are the same (experimentally determined in our model evaluation). This model achieves within 3\% performance of the similarity transformer and is able to generate a storyboard 8.4 times faster than the similarity transformer. 
However, our participants rated accuracy as highly important, so we believe they would be okay to wait more time for a more accurate and clean report, and thus use the 6k similarity transformer in our current system. 

\begin{table}[]
\centering
\begin{tabular}{@{}lllllll@{}}
\midrule
                          & P        & R        & F1  & Acc. & T(s)    \\
\textbf{Feiz Dataset - Baseline} & & & \\
SSim Transformer (FD) & 76.9\%   & 74.8\% & 75.8\% & 92\%  & - \\
SSim Transformer (1k) & 77.1\% & 87.1\% & 81.8\% & 94.5\% & - \\
\textbf{New Dataset - Our Work} & & & \\
SSim Transformer (1k)  & 82.2\%    & 94.1\%    & 87.7\% & 96.3\% & - \\
SSim Transformer (6k)   & 89.5\%    & 88.2\%    & \textbf{88.8\%}  & \textbf{96.9\%} & 42.2s  \\
SSim Bi-Encoder (Embedding) (6k)   & 91.1\%    & 81.2\%    & \textbf{85.9\%}  & \textbf{96.7\%} & 5s \\
\hline
\bottomrule
\end{tabular}
\caption{Performance results for the screen similarity (SSim) transformer and embedding based models (distance threshold $0.2$) demonstrating an improvement from our work of 13\% in F1 score from the baseline.}
\label{table:screen_sim_model_results}
\end{table}

\subsection{UI Element De-duplication Heuristics}
To evaluate the accuracy of our UI matching heuristics, we collected a dataset of 138k UI element correspondence labels across 25k same screen pairs from our annotated screen grouping dataset. 

\label{section:data_collection_annotation}
\subsubsection{Data Collection \& Annotation}
Half of the screen pairs (53.6\%) are very similar (MSE~\cite{meansquareerror} < 30). 4.4\% of pairs are the same screens with some content scrolled, while the remaining pairs have other content changes (e.g., added UI elements, removed UI elements, text content changes). Within each pair, our annotators consider two UI elements to be a match if they a) serve the same purpose (i.e., have the same functionality or convey the same information), b) are actionable and would lead to the same next screen in the app, and c) have the same grouping (e.g., an icon inside a container should be matched with the same icon and not the container). We include our annotation interface in the appendix. In the end, we found 17,913 (13.0\%) template UI element ($T_{ui}$) do not have any matching UI element in the new screen ($N_s$).

\begin{table}[]
\centering
\begin{tabular}{@{}lllllll@{}}
\midrule
                          & Precision        & Recall        & F1     & Time\\
Template Matching Only      & 87.5\%    & 96.3\%    & 91.7\%  & 2.57s \\
Exact Text Matching Added   & 89.4\%    & 95.4\%    & 92.3\%  & 1.11s \\
Fuzzy Text Matching Added   & 88.9\%    & 96.8\%    & 92.7\%  & 1.12s  \\
All Heuristics              & \textbf{97.7\%}    & \textbf{98.7\%}    & \textbf{98.2\%}   & \textbf{0.35s} \\
\bottomrule
\end{tabular}
\caption{Performance results for UI element matching\protect\footnotemark, reported by the subset of the matching heuristics applied.}
\label{table:fingerprinting_results}
\end{table}
\footnotetext{The average matching time for each template UI element was measured on a Macbook Pro with 2.4 GHz 8-Core Intel i9 / 32G memory}

We evaluated these heuristics on the 138k UI element matching annotations. To improve our heuristics, we examined and updated them on a small dataset of 991 screen pairs (5,420 template UIs), and then evaluated the heuristics on our full dataset. We report precision, recall, and f-1 score metrics using the following definitions: 

\textbf{True Positive:} The match found by our heuristics and the match found by the annotators are the same UI element.

\textbf{True Negative:} Our heuristics did not find a match and the annotators did not find a match. 

\textbf{False Positive:} Our heuristics found a match but the annotators did not, or the match found by our heuristics and the match found by the annotators are NOT the same.

\textbf{False Negative:} Our heuristics did not find a match but the annotators found a match. 

The performance on our full dataset of 138k UI element correspondences of our full set of heuristics, along with some variations,  can be found in Table ~\ref{table:fingerprinting_results}. First, we tried using image template matching alone to find the best match, which is quite slow as template matching must be run on each pair of UI elements. This baseline method achieves 91.7\% F1-score, in part because many non-Icon UI elements are not correctly matched. Next, we improve on the template matching method by adding exact text matching, which improves the F1-score by 0.6\%  and doubles the speed. Finally, we add fuzzy text matching to the previous methods, and the F1-score is further improved by 0.4\%. Fuzzy text matching tolerates small mistakes introduced by OCR imperfections, but may also create error when there are Text elements with small differences (see appendix). None of these baselines approach our full set of heuristics, which achieved a F1-score of 98.2\% and dramatically improves speed as more UI pairs can avoid template matching.

We also examined the performance of our heuristics on three common cases of screen similarity. For screens with very few differences, it is often possible to match the UI element simply by location. Our method works almost perfectly on these types of screen pairs (99.2\% Precision, 99.5\% Recall, 99.4\% F1-Score). It also works well on scrolled screen pairs (91.7\% Precision, 96.8\% Recall, 94.2\% F1-Score), and screen pairs with other content changes (95.1\% Precision, 97.0\% Recall, 96.1\% F1-Score). We examined the failure cases and share common patterns in the appendix.

\begin{figure}[t]
    \centering
    \includegraphics[width=\linewidth]{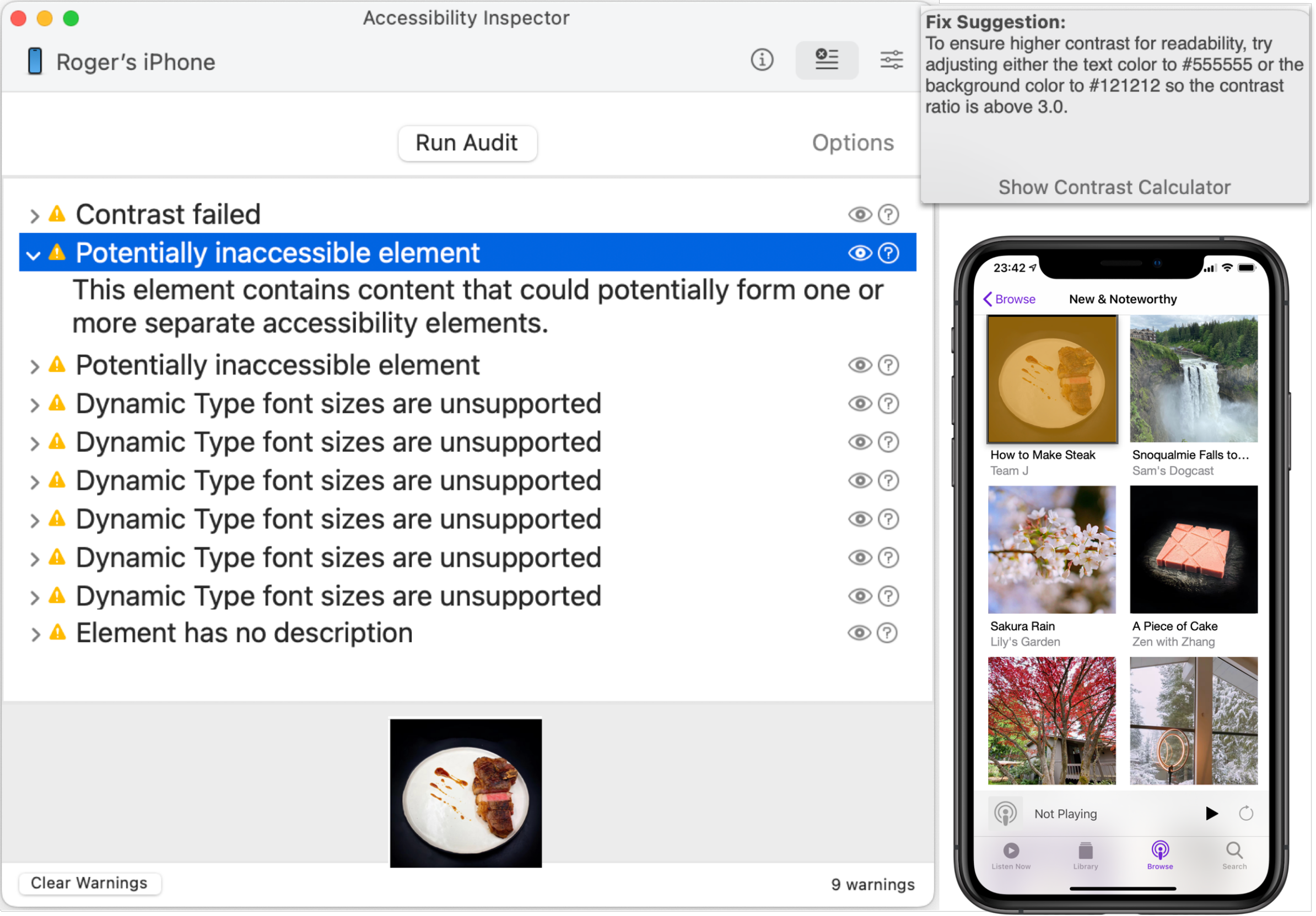}
    \caption{Accessibility Inspector in single screen mode (SS) with audit results for an iPhone.}
    \Description{The figure shows an interface on the left containing a list of detected accessibility isssues for the device on the right. The table contains rows, underneath which contains a description of the issue found. The tool also provides fix suggestions which indicate what should be done about the issue. }
    \label{fig:axi}
\end{figure}

\section{User Study}
To evaluate our report generation system, we conducted a study to better understand how the system can impact QA testers and developers' ability to gain awareness and prioritize issues to be fixed across an app. We evaluated the following research questions: 
\begin{itemize} 
    \item How does the mode of accessibility scanning impact users' creation of quick accessibility reports?
    \item How do users perceive the quality of our automatically generated accessibility reports?
    \item How should accessibility auditing tools support prioritization and quick discovery?
    \item How can accessibility reports fit into participants' workflows?
\end{itemize}

\subsection{Participants}
We recruited 19 (5 F, 13 M) participants across a large technology company to take part in the study, across varied roles including software engineer (9), QA or Automation engineer (7), accessibility evangelists (1), and managers (2). Participants mean self-rated expertise in iOS app development was \red{3.31 (Med: 4, Std: 1.6)} and in accessibility testing was \red{3.8 (Med: 4, Std: 1.01)} where the rating scale consisted of: 1 - No experience, 2 - Beginner, 3  - Advanced Beginner, 4 -  Intermediate, 5 - Expert. 18 participants were sighted and some used varying degrees of magnifications features, and 1 participant used a screen reader. 

We report the screen reader user's results separately as they were only able to complete two tasks during the allotted time for the session; however, they provided valuable feedback that may make our system more accessible in the future to developers who are also screen-reader users. 

\subsection{Procedure}
During each study session, we began by asking participants to describe their prior experience in using accessibility testing and reporting tools. Then, participants completed 3 tasks in a counterbalanced order. For each task, we instructed the participants to conduct an accessibility audit of 3 different apps using 3 different sets of tools. The three sets of tools included:  
\begin{itemize}
\label{itemize:conditions}
    \item \textit{Single screen accessibility inspector (SS)} - This tool, showin in Figure~\ref{fig:axi}, supports single screen auditing with a button "Run Audit" which when clicked returns a list of possible issues  for a screen across 28 possible issue types. Clicking "Run Audit" clears the results from the prior audit.
    \item \textit{Multi screen accessibility inspector (MS)} - This tool adds a history to the results of the previous mode where the interface keeps audited results for each new screen, and combines them with the results for prior audits of the same screen. The mode additionally adds a header on the results for each screen to report the number of issues found in each category for that screen (e.g., 3 Element Description, 2 Contrast). With this mode, our goal was to introduce features aimed to help users with summarization and prioritization of issues across screens, while keeping manual auditing constant. 
    \item \textit{App crawler with generated report (AC)} - This tool provides a pre-generated accessibility report, generated by our system's app crawler, with results hosted in a web page for participants to examine (see Figure~\ref{fig:report_ui} for an example report). 
\end{itemize}

For each accessibility auditing task, we asked the participants to complete a QA task to briefly find and prioritize 3-5 key issues across the app that would hypothetically be sent to a development team for that app to fix. We instructed participants to add context to the report to help the developers reproduce or interpret the issue, and optionally add screenshots, but that was not required and no participants opted to do so during the study. We gave participants 6 minutes to complete each auditing task. The reason for making this a short task was to see how our tools could be used to conduct quick app accessibility audits, and to keep the total session length under 45 minutes while leaving time for follow up interviews. Participants completed 3 such app auditing tasks where we assigned one tool (as  listed in ~\ref{itemize:conditions}) and one app to audit for each task. We counterbalanced the order participants used each tool and the apps being audited using a Latin square ordering across participants. We selected three publicly available apps (which we refer to as app A, B, and C) from the top 100 apps for each three categories in the App Store initially at random; and then to have apps with a roughly equivalent complexity and number of accessibility issues. We ran our app crawler on each app to produce a report (example seen in Figure~\ref{fig:report_ui}). The number of screens in each app was 39, 63, and 58 for apps A, B, and C, respectively while the number of total warnings surfaced in the accessibility report was 444, 596, and 522 respectively. 

As we found during Study 1, participants typically will also manually validate issues found by accessibility scanning tools. Therefore, we also allowed participants to manually validate issues flagged by each tool using a locally attached iPhone. 

After participants completed the 3 tasks, we interviewed them and had them complete a follow-up survey to compare and contrast their experiences using each available tool for the task. We recorded audio for each session and took notes on how many screens each participant audited for the SS and MS tasks. We saved each participant's report of the issues found for each app in a document. We conducted a qualitative thematic analysis~\cite{guest2011applied} of the interview results, and we rated the quality of reported issues through a rubric to measure the specificity, scope, and importantness of the issues reported by participants in their summaries. 

\subsection{Results}
In this section, we summarize the results per each research question we examined in the study. 

\subsubsection{How does the mode of accessibility scanning impact users' creation of quick accessibility reports?}
Overall, 13 participants preferred AC (app crawler) the most, while 3 preferred MS (Multi-screen accessibility inspector) and 1 preferred SS (Single-screen accessibility inspector). Two participants did not prefer any one mode, mentioning finding useful features with both MS and AC modes. One participant preferred manual testing over all modes as they thought they would be faster to move through the issues by hand. 
Participants gave reasons such as ``giving a more holistic overview'', ``saving manual effort'', ``removing the dependency on the Xcode and the device'', being ``more sharable'', and ``reducing friction and context switching''. 

Conversely, the majority of participants (13) had the least preference for SS giving reasons such as '``required more context switching'', ``more cognitive load'', and ``more manual effort'' to capture multiple screen audits. A few participants preferred MS the least while the remaining did not prefer one tool the least. Those participants mentioned seeing value in all three tools, or didn't see much difference between MS and SS for this task.

\textbf{\textit{Which tool helped participants create better reports?}}
Participants were overall more satisfied (using a 5-point Likert scale for satisfaction) with their reports created using AC (Mean: 4.05, Med: 4), compared to MS (Mean: 3.68, Med: 3) and SS (Mean: 3.35, Med: 3). The reports contained many dynamic type issues, missing labels, poor contrast, and small target size, but some issues participants found are not detectable by our system and participants found them through manual testing. 

As satisfaction can be subjective, we also evaluated the contents of the reports. Two authors rated the issues listed in each report for each app and condition using a rubric consisting of three categories - \textit{Specific}, \textit{High Level}, and \textit{Important}. For \textit{Specific}, we rated each issue on how easily we could deduce the UI elements or screens impacted by the issue. For high level, we rated each issue on whether it applied to a single element (1), a single screen (2), or multiple screens across the app (3). For \textit{Important}, we rated the severity of the issue based on whether it would block any major usage of the app for key accessibility features (e.g., Voice Over, large text) on a scale of 1 to 3. Through this rubric evaluation, we sought to understand whether any particular condition helped the participants in creating reports that covered a wider scope of important issues across the apps. 

Each rater rated all 163 issues listed by the 18 participants along these dimensions. After rating, if any rating differed by at least two for a category, the raters discussed the grading and resolved disagreements if possible. Ultimately, the raters achieved an IRR (Percent Agreement) of 0.80 and IRR for ratings differing by 1 or less was 0.99.  

Overall, the mean ratings per listed issue across all three modes were very similar for \textit{Specific} -- AC was 2.38 (Med: 2.5, Std: 0.61) while MS was 2.37 (Med: 2.5, Std: 0.65) and SS was 2.32 (Med: 2.5, Std 0.63). However, there was a larger difference for \textit{High-Level} -- The ratings for AC (Mean: 2.12, Med: 2, Std: 0.79) were 9\% higher than MS (Mean: 1.95, Med: 2, Std: 0.72) and 16\% higher than SS (Mean: 1.83, Med: 1.75, Std: 0.78). Finally for \textit{Important}, the ratings for AC (Mean: 2.23, Med: 2.5, Std: 0.7) were 13.3\% more than MS (Mean: 1.97, Med: 2, Std: 0.76) and SS (Mean: 1.93, Med: 2, 0.74). While there were differences between the means between tools, we found these differences not significant using the Aligned Rank Transform~\cite{wobbrock2011aligned} for non-parametric data with \textit{Tool}, \textit{Experience}, and \textit{App} as factors.

\subsubsection{How do users perceive the quality of the generated report?}
Overall, participants thought the app crawler report was "clean" (Likert scale, 5: Very clean, 1: Very messy) with a median rating of 4 (Mean: 3.85) and rated the screen grouping quality (Likert scale, 5: Very accurate, 1: Not accurate) as accurate (Med: 4, Mean: 3.65). 

When rating screen grouping quality, we asked participants to open the AC report and showed them a few examples of grouped screens across the report, as most participants did not notice this feature right away. Participants were also not super familiar with the structure of each app and as such their responses to this question may not be as grounded as they might if they were the original developers of the apps. 

\subsubsection{How should accessibility auditing tools support prioritization and quick discovery?}
We additionally examined whether any particular tool or features of any tool helped in prioritizing and summarizing issues by asking the following questions: 
\begin{itemize}
    \item Which tool helped you discover the issues most quickly? 
    \item Which tool helped you most to find the most common issues?
    \item Which tool helped you most to prioritize the \textit{most important} issues?
\end{itemize}

\textbf{\textit{Which tool helped you discover the issues most quickly?}}\\
Perhaps unsurprisingly, the app crawler was rated as the most helpful tool in discovering the issues most quickly (14 participants). Participants were also able to audit an order of magnitude more screens across the app using the crawler (A: 39, B: 63, and C: 58)  vs the other modes. Using the SS tool, participants audited 3.9 screens on average (Med: 3, Std: 2.78) and with MS they audited 4.5 screens on average (Med: 4, Std: 2.91). Participants (n=12) noted the manual modes could be a bottleneck, time consuming, and tedious, especially with limited time for auditing  allocated during the study.
\quotateblock{P2}{(SS) took too much time and i wouldn't have been to even get all the views up to look through them in time ... It was a bottleneck to have to keep loading and running the tool on each view.}

Another aspect of effort saved on the AC mode participants noted (n=5) was more hypothetical future use cases in that the report would ultimately ``reduce friction'' by removing dependencies on setting up a device for testing. 
\quotateblock{P1}{I don't have to worry about being on the latest version or any incompatibilities ... I can just go to the website with all the data, interact with it, file my bugs and then go from there.}

Conversely, participants noted that using SS to create the report required more context switching compared to AC and MS. In MS, the UI adds a ``history'' of captured audit results which supported tracking for participants and reduced context switching, making it easier to identify patterns in the data. 
\quotateblock{P9}{I guess in the multiple, like, for example, if I'm seeing his area is too small and I'm seeing it multiple times. I kind of like, okay, this is maybe like a main issue on this.}

Although participants found the history added in MS useful for finding patterns, some participants found MS more confusing and would use SS to target a specific screen of interest. 
\quotateblock{P7}{(MS) was high on the level of complexity for like, the really primitive that the accessibility inspector. Usually if I'm trying to, um, audit like a specific thing, I would just use the single screen accessibility inspector, uh,  to look at the one particular screen of interest.}

Participants (n=6) also felt they were able to get more coverage and more information across the app to make decisions with the app crawler as compared to the other modes: 
\quotateblock{P13}{But, yeah, I guess for even for, for contrast issues and stuff, like, the last tool is really great, because I felt confident that we got like, a really good look at the entire app.}

While more information was seen as helpful in getting a better look at the app, six participants noted that it could also be overwhelming to look at, which might be discouraging or make it difficult to know where to start. 
\quotateblock{P2}{I can see that it being like too overwhelming because then there's like, oh, now we have a, I got a huge list of, like, you know, 500 accessibility issues. Which ones do we start with?}

\textbf{\textit{Which tool helped you most in finding the most common issues?}}\\
The majority of participants reported that AC helped them most to find the common issues (14 participants) across the app. Participants (n=8) noted that AC assisted them in finding the most common issues by providing a summary and counts for each category across the app. Participants noted that the counts and summary helped them to spot more prevalent issues across the app, like missing support for Dynamic Type, a pervasive issue among the apps tested in our study. 
\quotateblock{P12}{You can click here if you want to know the count, but I'm gonna give you the full list and and after while you're going to be able to just to answer this and be like yeah, we got a big problem with the Dynamic Type. I just thought that was such a great delivery.}

Counts of issues were also useful in the MS mode, as two participants mentioned this mode helped them most, evidenced by P15 spotting issues popping up ``It seems like dynamic type. Font sizes are unsupported is popping up across multiple audits.''

\textbf{\textit{Which tool helped you most to prioritize the \textit{most important} issues?}}\\
The results were more mixed with 7 participants choosing the app crawler while the remaining chose MS (6), SS (1) and Multiple Tools (2) and None of the above (2). Participants that chose either MS, SS or Multiple Tools mentioned that because the AC randomly explored the app, they had no control over whether important user tasks were covered in the app. In the other modes, they could quickly review key user scenarios by controlling which screens were audited. \quotateblock{P14}{If we can, if we can attach a process to this, and then I'm just going through, let's say, or an ordering flow, and then generates the report for this ordering flow ... let's just say, hypothetically, my team is in charge of building the order flow. I'm not going to care about accessibility of the other screens}. This suggests that participants may benefit from having a mode that supports both \textit{automated exploration} and \textit{control} over which user scenarios and tasks are explored by the app crawler.  

\textbf{\textit{Prioritization and Discovery: Strengths and Potential Improvements}}
One key theme in our study was that a high level report across the app gave participants new capabilities and benefits compared to manual auditing modes. Participants (n=14) mentioned that the overview, summary, and total counts of the AC report helped them strategize and prioritize which issues to fix. Total counts helped P19 ``I'm just going through and kind of looking at I'm trying to look at where are the warnings are the highest?'' to organize how they looked through the report. By clicking on categories in the summary tab, participants could view the total counts of issues and visualize all impacted screens at once. This helped them discover higher level patterns of issues. 
\quotateblock{P7}{The tool is reporting that we have, you know, uh, far and away, more large type issues than anything else. Maybe that's where we need to focus our attention maybe not if they're like, small, minor things.}
While participants in general felt that the AC report helped them prioritize and report the issues across the apps in the task more easily than the other modes, they gave several creative and insightful suggestions about how to make the reports more useful and more interpretable for future developers. Several participants in the study had questions about particular issues and what they mean, suggesting that the fix suggestions provided in the AC report and the SS and MS tools could be clearer, provide more detail, or ``link to additional resources'' (P12).  

Several participants also found the report initially to be ``overwhelming'' and noted that it might discourage developers to see a big number of issues. However, participants saw the potential for the system to better help in prioritization through varied suggestions such as: 
\begin{itemize}
    \item Assigning priority or severity ratings to each issue category. 
    \quotateblock{P12}{The problem is that no one knows what the importance and priority of these fixes are so if engineering came back and said, no, we will fix 44 of these 444 issues you say what is the top 10 most critical fixes that we need here we judge all fixes to be identical. Level 1, issue, level 2, issue, level 3, issue}
    \item Statistics or emphasizing the magnitude of impact on a disability population. 
    \quotateblock{P14}{Maybe we can actually empower people to get some, some stuff done. I mean, saying there's a 450 dynamic type issues versus saying oh, there's X number of 100,000 people that won't be able to use your your tool. This many of your bugs affect people with low vision, this many people effect users of screen readers.}
    \item Importance or frequency of usage of the impacted UI elements. 
    \quotateblock{P14}{If we only had, let's say, half an hour to to triage all these issues, let's do all the highly visible, highly interacted with it. Yeah. First, and then let's commit to fixing this thing, like, you know, this sprint or next sprint.}
    \item Scope of the issue across the app and reporting higher level insights and themes. 
    \quotateblock{P13}{Like, if it went through every screen, it was like, okay, like, 90\% of the elements didn't support dynamic type then probably like limited dynamic type for support altogether. Yeah. So, it'd be nice to just, you know, pull up these into one area for dynamic type.}
\end{itemize}

\subsubsection{How can accessibility reports fit into participants' workflows?}
Throughout their responses to multiple questions in the study, participants hypothesized about workflows where MS and AC modes might be useful.  The AC report was seen as particularly useful for QA and reporting use cases (n=9), where QA testers might want to track the accessibility of the app over time or compute stats and trends.
\quotateblock{P16}{Like, is there a way to do a checklist? So once you make fixes, can you show, like, how it's been fixed over time or changing over time, or something like that?}

Related to this, participants (n=4) wanted to integrate AC into their continuous integration workflows, which would enable them to run the reports on a regular basis, across multiple devices and settings (e.g., dark and light mode), or for testing accessibility across multiple languages. 

Supporting triaging and marking issues as ignored over time was also noted as an important feature in long term use, as participants mentioned they might be likely to file bugs or find issues in the report that they would mark as ``won't fix'' or ``minor issues''. These issues should then be filtered out of future reports automatically.  

Aside from it's usefulness in QA and continuous integration, several participants (n=7, mostly software engineers) desired for the tool to be closer to Xcode where they typically develop their apps. As opposed to a dynamic accessibility scanning approach which runs on a built app like our system, some participants desired for this auditing to be done as they were building the app or on specific screens when possible, similar to the functionality of existing accessibility linters~\cite{googleAndroidLint} that statically analyze code for accessibility issues.  
\quotateblock{P5}{
Like, when I do my work, I have a workflow, I would like to keep, productive, 
I have to minimize the switch from tool to another. Having the accessibility inspector, not part of Xcode. It becomes like a like, how to describe that. Uh, it becomes like a detour.}

\subsection{Screen Reader User Feedback}
One participant in our study is a screen reader user. They were only able to complete two tasks during the 45-minute session, so we report the results separately to reveal both benefits our system gave this participant over other auditing tools and areas of future improvement. First, a key benefit this participant noted was that for very inaccessible apps, our system would let them generate a more complete report compared to manual scanning tools that require use of the VoiceOver screen reader to navigate the app to each screen for auditing. Unexposed elements in very inaccessible apps might cause them to miss auditing key areas of the apps they would be unable to navigate. 

However, as our system cannot yet provide automatic alt-text to describe each screen at a high level, the participant struggled to establish context to navigate the report to determine which issues belonged to which screens in the app, which would be necessary to include when filing a bug report. In the future, we may be able to address these limitations with recent UI understanding technologies~\cite{wang2021screen2words,wu2021screenparsing}.

\section{Experience Reports}
In our user study, participants audited accessibility issues for apps they did not own themselves. Thus, we also gathered experience reports from app developers within our organization to understand how they might leverage our manual auditing mode (MS in our user study) and app crawler generated reports (AC). We generated AC reports for 5 app developers to evaluate their own apps' accessibility and gave them our MS tool to use on their own apps (see the multi-screen version inspector in Figure~\ref{fig:axi}). The apps they worked on were internal apps used within our organization for bug filing, device and build management, and sample apps from documentation. Each developer we interviewed did have QA in place for accessibility testing, although the majority of them (4) lacked automation tests and primarily relied on manual testing. 

\textbf{\textit{Issues Found}}
Each developer looked through the AC report and noted any issues they found for which they might file a bug. The mean number of screens in these reports among the five apps was 30.4. Since the developers were already highly aware of accessibility, most of their apps did not have pervasive accessibility issues. However, each located screens in their app lacking Dynamic Type support, and at least one UI element with an incorrect or missing accessibility label. The developers typically manually verified these issues themselves outside of our tool, which may be due to lack of trust in prior tools which provided false positives. However, a common thread of feedback on both MS and AC tools was that the reports contained issues (e.g., low contrast, target size) flagged on system controls or system provided dialogs or screens they had no control over. A challenge for future versions of our system is to filter these out or report them separately. 

\textbf{\textit{Workflows}} Three developers noted some issues in the report they had decided to not fix, or had alternative solutions for after discussing with accessibility QA. They would ignore these issues if this tool was integrated into their workflows. All developers expressed excitement about the AC reports, requested access to it, and  envisioned using it in their workflows.   Developers thought that MS was better than prior versions of this tool (single screen mode), but would prefer to use the AC report if available for CI or through a command line tool. 

\textbf{\textit{Coverage}} Three developers noted that the AC report contained all the key screens they were able to think of while two found that it missed capturing some key screens and areas of their apps. Using the MS tool, those developers audited those screens themselves but did not find any additional issues they would file in those two cases. In future versions of our system, we intend to improve our app crawlers to obtain more complete coverage.

\section{Discussion \& Future Work}
Overall, our system received positive feedback from both our internal stakeholders and participants in our studies, indicating that accessibility reports helped them summarize and prioritizing issues. The two key technical contributions of this work, screen grouping and UI element matching, also achieved high accuracy. We plan to continue improving these components of our system in future work while also addressing the improvements noted by our stakeholders and study participants.

In this work, we primarily focused on evaluating our models and the overall idea and features of the system over the design of the report itself and how it conveys information. However, our study revealed insights around how future systems might surface accessibility issues in reports. Some of these insights can directly leverage our introduced models, to report the scope of issues, for example. Our participants were also relatively familiar with app accessibility. Future work should evaluate and explore how to best present and prioritize summarized issues to make them easily understandable to developers not as familiar with accessibility as our study participants.

While our screen grouping model achieves high accuracy, our data still contains many annotation errors which we would like to resolve to improve the model further. We also plan to collect additional crawl data for screen variations the model has difficulty predicting (e.g., scrolled screens, keyboard open / closed). After improving these models, we will continue to evaluate our system with users to study the impact of accuracy improvements. For future work, we will  apply these models in other contexts beyond accessibility report generation such as UI testing, design, and record \& replay systems. 

Additionally, we plan to further explore methods of accessibility report data collection. Our prototype supports both manual capture and random crawling. While we used the random crawler to produce reports for our study, we have not evaluated crawler coverage. We plan to further develop and leverage improved app crawling in future work, based on feedback from our user studies indicating that they would like more control over reporting specific application flows that are relevant to them.

While our current prototype leverages the Accessibility Inspector~\cite{accessibilityInspector}, this tool does not currently cover every available accessibility check. It is unlikely that automated tools will ever find all types of accessibility issues~\cite{mateus2020accessibility}, and it will still be important to leverage other types of accessibility testing (e.g., manual test suites, automated tests with accessibility services like VoiceOver) in addition to auto-generated reports. It is also important to continue to involve users with accessibility needs in the testing process. However, as our study participants and internal stakeholders note, this tool can be a valuable complement to their current testing. We also plan to explore additional ways machine learning can be used to detect and report additional issues (e.g., grouping, navigation order) to make the report more informative. 

To better contribute to reproducibility of our research, we hope to release our datasets for use in the research community at some point in the future.

\section{Conclusion}
In this paper, we presented a system to generate accessibility reports for mobile apps from a variety of input sources. Our technical evaluations of our models and user evaluations of our system demonstrate this approach is promising and can provide value as a tool in the accessibility testing process for mobile apps. Our screen grouping model and UI element matching methods may also have implications in a number of UI testing and interactive applications beyond report generation. Going forward, we will continue improving the accuracy of our methods and usability of our reports to enable developers to quickly comprehend key areas where their app's accessibility should be improved.

\bibliographystyle{ACM-Reference-Format}
\bibliography{references}

\newpage
\appendix
\section{Appendix}

\subsection{UI Element Matching}
\begin{figure}
    \centering
    \includegraphics[width=\linewidth]{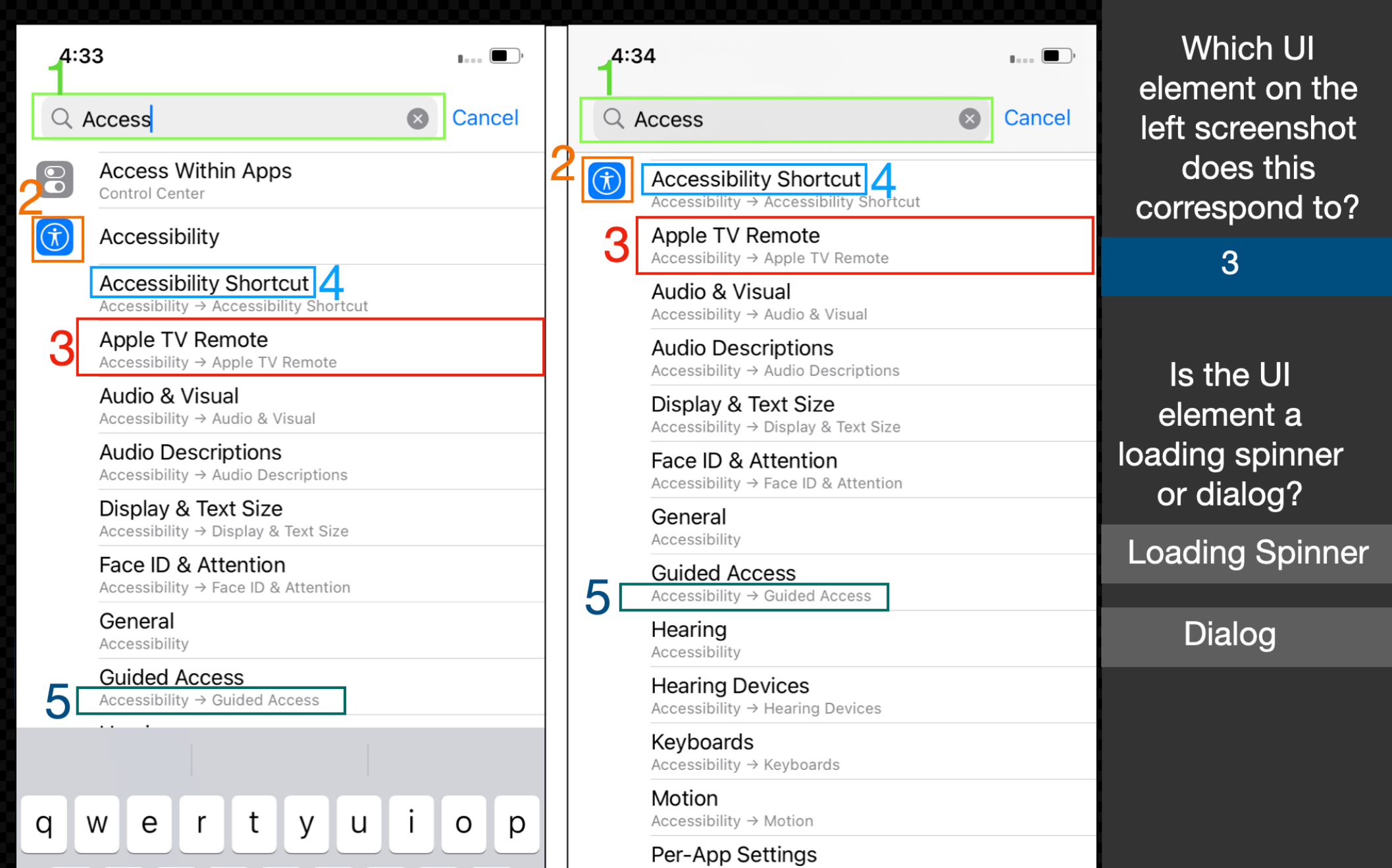}
    \caption{The interface crowd workers used to label the evaluation set for UI element matching. Each task displayed a left and right screenshot. Crowd workers drew boxes and labeled matching UI elements on the right screenshot, using the left as a baseline}
    \label{fig:ui_matching_interface}
\end{figure}

To collect our UI element matching evaluation dataset, crowdworkers used the labeling interface pictured in Figure~\ref{fig:ui_matching_interface}. Figure~\ref{fig:ui_matching_interface} shows the interface we used to collect these labels. Each annotation task contains two screens. The left screen ($T_s$) shows 6-8 highlighted template UI elements with corresponding numbers of UI elements to be matched in the right screen ($N_s$). We provided several examples of UI element matching to supplement the definitions of UI element matching.

\begin{figure*}[b]
    \centering
    \includegraphics[width=\textwidth]{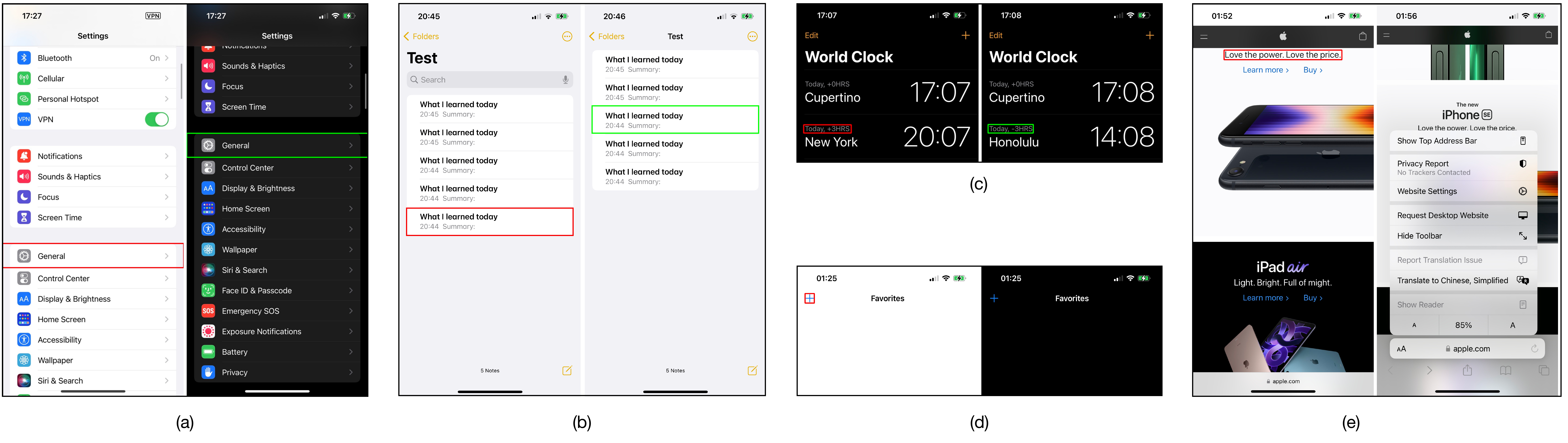}
    \caption{Examples of True Positive (left) and False Positive (right) of UI element matching heuristics. The red box indicates target UI on original screen, while the green box indicates the matched UI on the new screen. Failure examples of UI element matching heuristics.}
    \label{fig:ui_element_matching_failure}
\end{figure*}

\subsubsection{Failure Cases}
To understand common patterns for future improvements, we examined failure cases on our dataset. 

\textbf{False Positive}
\begin{itemize}
    \item When multiple UI elements are the same as our template UI (e.g., notes in Figure ~\ref{fig:ui_element_matching_failure}(b)), our method may find a wrong match as the similarity scores are close. It suggests our similarity score may consider the distance to template UI.
    \item When the new screen does not contain template Text, our heuristics sometimes will find a very similar Text as the best match. As seen in Figure ~\ref{fig:ui_element_matching_failure}(c), our method finds "Today -3HRS" as the best match when fuzzy text matching ignores one character difference. Fuzzy text matching might use a higher threshold in future applications. 
\end{itemize}

\textbf{False Negative}
\begin{itemize}
    \item When the background color of icon changes (e.g., light / dark mode in Figure ~\ref{fig:ui_element_matching_failure}(d)), our image template matching method has a hard time finding the icon, as most of the pixels inside icon bounding box changed. During template matching, future method should extract background color of icons and only consider the stroke color of icons.
    \item When the matching UI element is partially occluded (as shown in Figure ~\ref{fig:ui_element_matching_failure}(e)), human annotators often perform better than our method.
\end{itemize}

\subsection{Storyboard Generation}
\begin{figure*}[t]
    \centering
    \includegraphics[width=\textwidth]{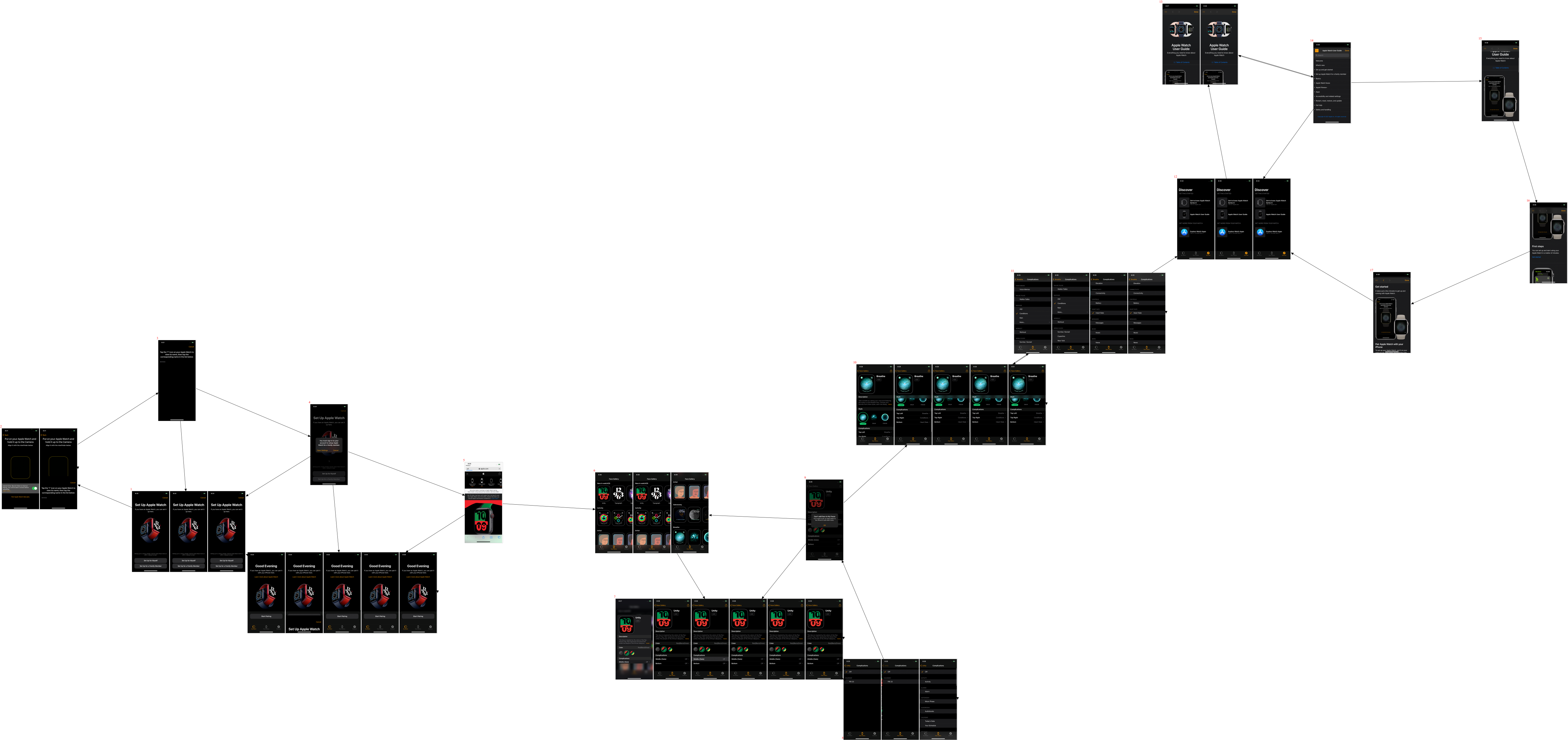}
    \caption{Similarity Transformer Generated Storyboard Example}
    \label{fig:storyboard_example_1_b}
\end{figure*}

\begin{figure*}
    \centering
    \includegraphics[width=\textwidth]{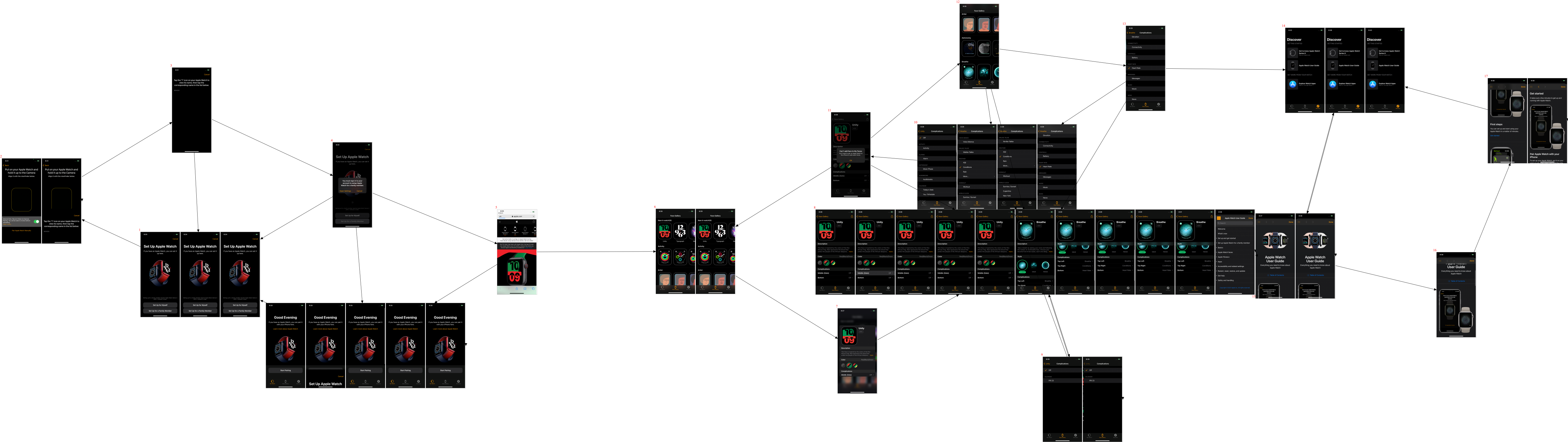}
    \caption{Bi-Encoder Storyboard Example}
    \label{fig:storyboard_example_2_sim}
\end{figure*}

We provide the full example storyboard for Figure~\ref{fig:storyboard_examples} generated by the similarity transformer~\ref{fig:storyboard_example_1_b} and bi-encoder models~\ref{fig:storyboard_example_2_sim}.

\begin{figure*}
    \centering
    \includegraphics[height=\textwidth]{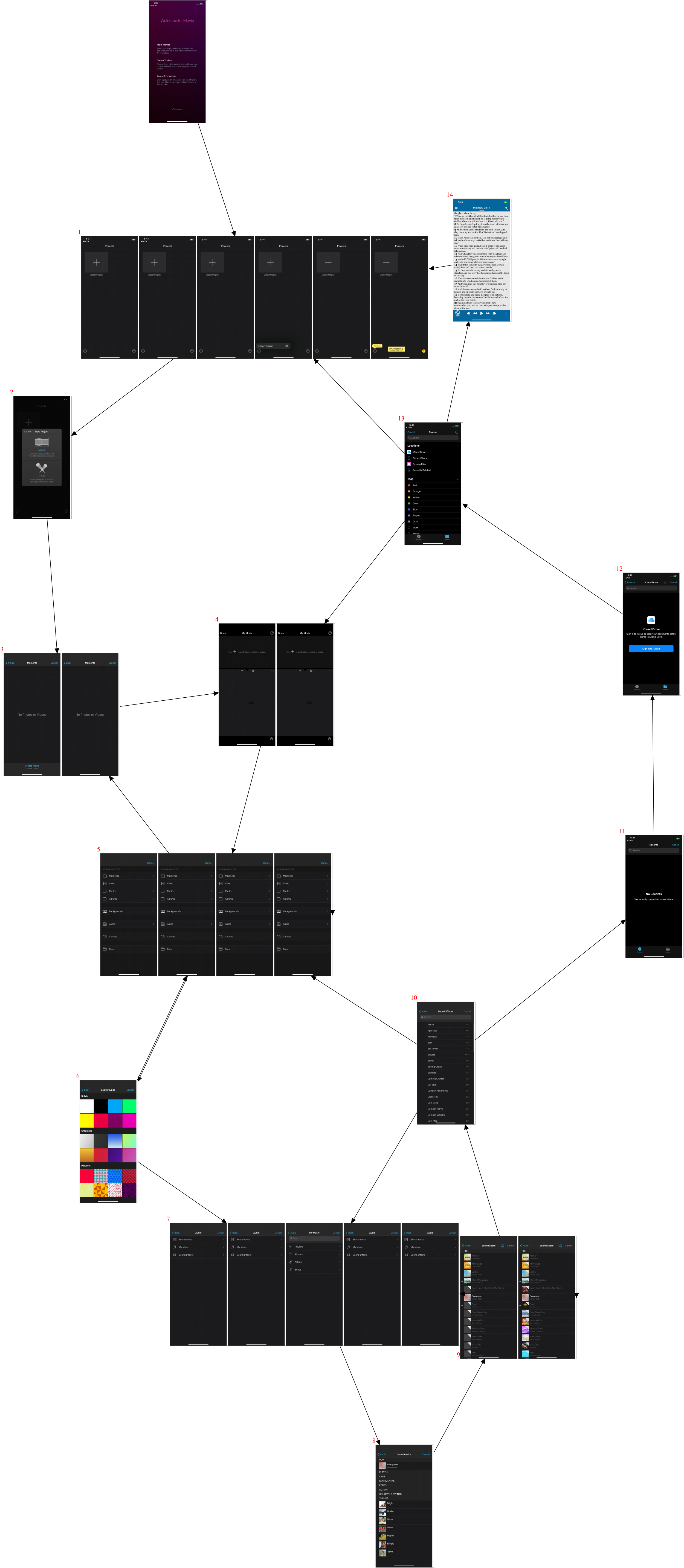}
    \caption{Similarity Transformer Storyboard Example}
    \label{fig:storyboard_example_2_b}
\end{figure*}

\begin{figure*}
    \centering
    \includegraphics[height=\textwidth]{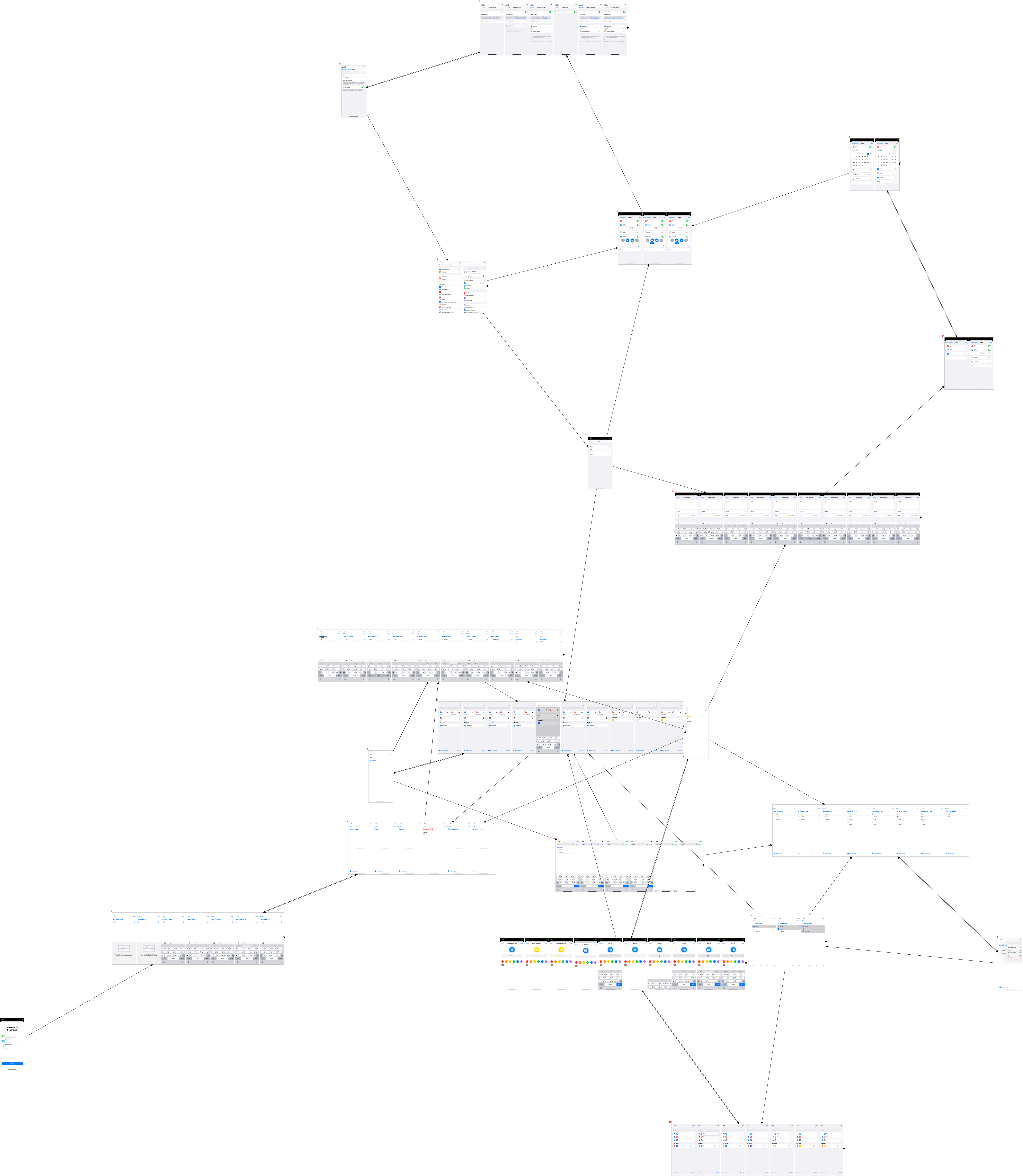}
    \caption{Similarity Transformer Storyboard Example}
    \label{fig:storyboard_example_3}
\end{figure*}

\begin{figure*}
    \centering
    \includegraphics[width=\textwidth]{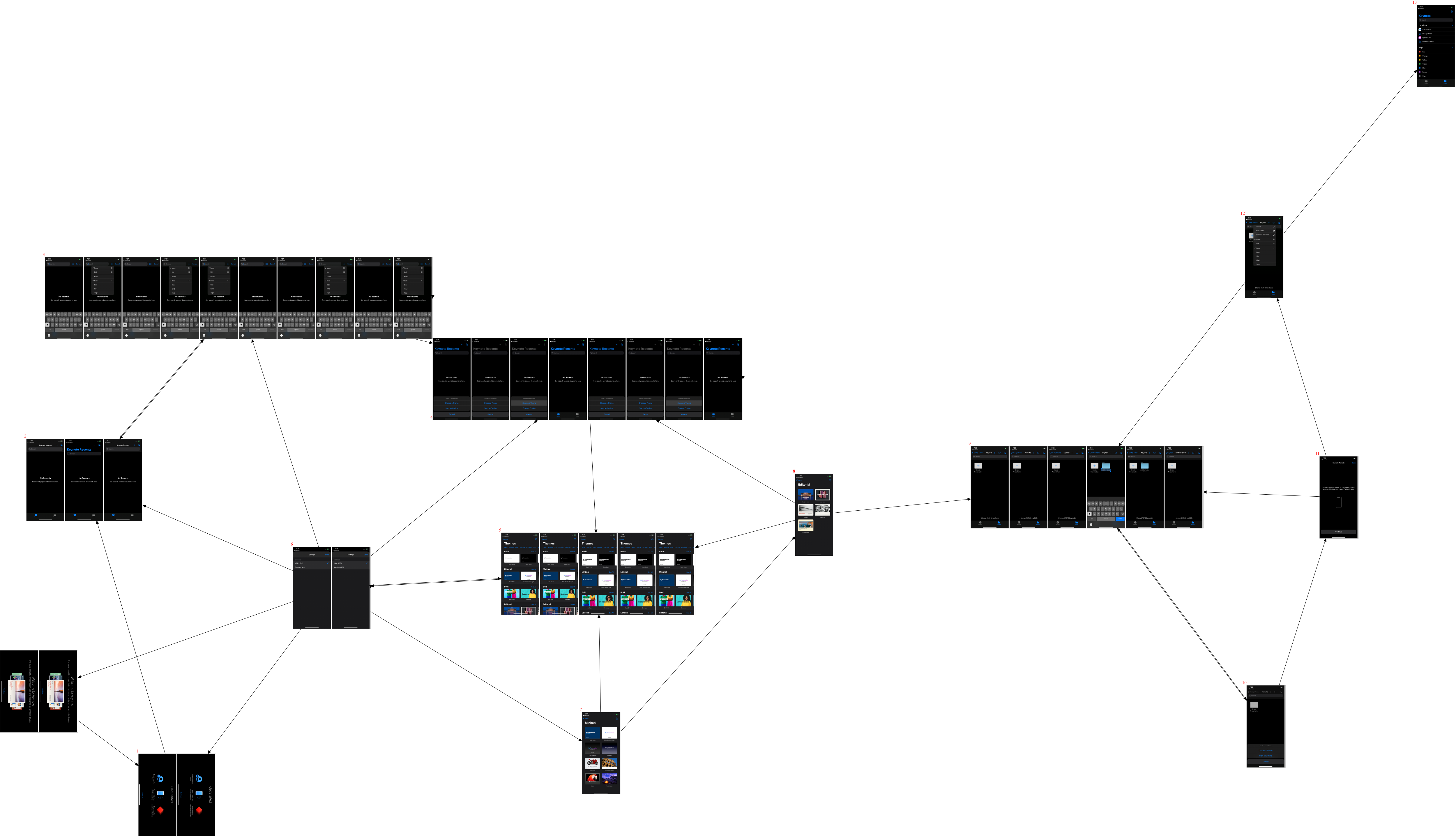}
    \caption{Similarity Transformer Storyboard Example}
    \label{fig:storyboard_example_4}
\end{figure*}

\begin{figure*}
    \centering
    \includegraphics[width=\textwidth]{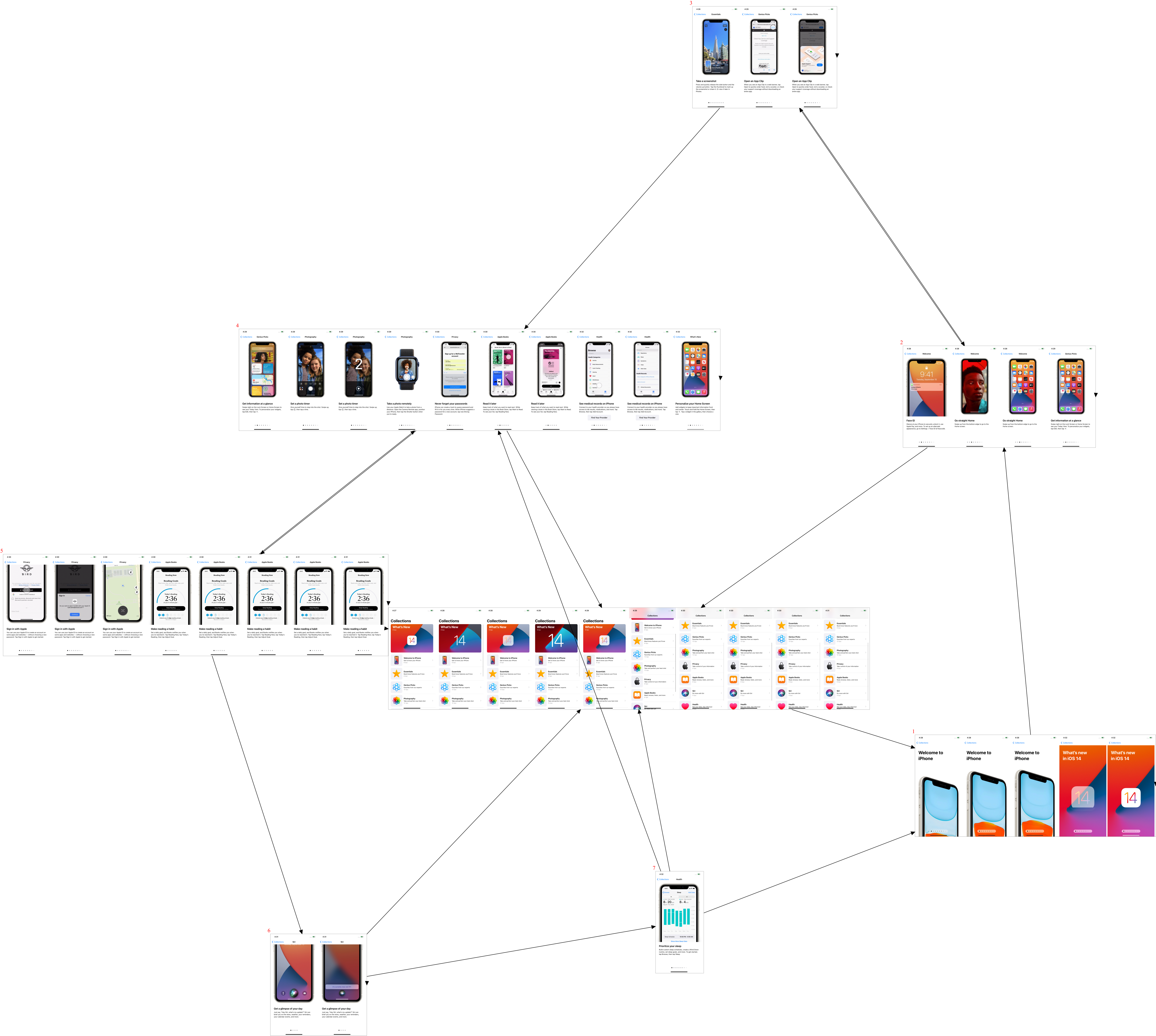}
    \caption{Similarity Transformer Storyboard Example}
    \label{fig:storyboard_example_5}
\end{figure*}

\end{document}